\documentclass[a4paper]{article}
\usepackage[english]{babel} 
\usepackage{graphicx}
\usepackage{hyperref}
\hypersetup{setpagesize= false,colorlinks = true, urlcolor = blue, linkcolor = black, citecolor = black
}
\usepackage{authblk}
\usepackage{geometry}
\usepackage{setspace}
\usepackage{verbatim}
\usepackage[utf8]{inputenc}
\usepackage{amsmath}
\usepackage{amssymb}
\usepackage{gensymb}
\usepackage{verbatim} 
\usepackage[english]{babel} 
\usepackage{minted}
\usepackage{amsfonts}
\usepackage{cancel}
\usepackage{indentfirst}
\usepackage{physics}

\usepackage{dsfont}

\newcommand{\keywordsenglishname}{Keywords}

\renewenvironment{abstract}{%
      \begin{center}
	\begin{minipage}{14cm}
	{\textbf{\abstractname:}}
}{
      \end{minipage}
	\end{center}
}
\newenvironment{abstractinenglish}{
      \def\abstractname{\abstractinenglishname}
	\begin{abstract}
}{
      \end{abstract}
}

\newenvironment{keywordsenglish}{
\def\abstractname{\emph{\keywordsenglishname}}
	\begin{abstract}
}{
\end{abstract}
}

\title {Exploring the nature of gravity with quantum \\ information methods\thanks{This manuscript is based on an article that was published in Brazilian portuguese language in the
journal Revista Brasileira de Ensino de Física, vol. 47, suppl. 3, e20250371 (2025), https://doi.org/10.1590/1806-9126-RBEF-2025-0371.}}
\author{Bruna Sahdo \href{https://orcid.org/0000-0000-0000-0000}
{\includegraphics[scale=0.04]{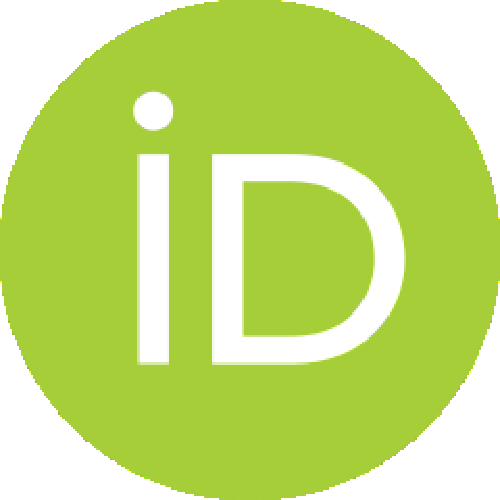}}$^{1,2}$}
\author{Nat\'alia Salom\'e M\'oller \href{https://orcid.org/0000-0003-0949-9023}
{\includegraphics[scale=0.04]{orcidicon.eps}} \thanks{natalia.moller@savba.sk} $^3$ }
\affil{$^1$ University of Vienna, Faculty of Physics, Vienna, Austria.}
\affil{$^2$ Institute for Quantum Optics and Quantum Information,
Austrian Academy of Sciences, Vienna, Austria
}
\affil{$^3$ Research Center for Quantum Information, Institute of Physics, Slovak Academy of Sciences, Bratislava, Slovakia.}
\date{}

\usepackage{fancyhdr}
\begin{document}

\maketitle
\vspace{6pt}
\begin{abstractinenglish}
The aim of this article is to provide an introduction to the use of quantum information methods for investigating the interface between quantum theory and gravity. To this end, we discuss the basic principles of two current research streams that use this approach. The first one explores a phenomenon known as gravitationally induced entanglement, which aims to infer whether the gravitational field responsible for the interaction between two massive bodies must be quantized or not.
The second stream investigates causal structures, thereby providing indirect evidence that spacetime may exhibit non-classical behavior.
Before presenting these topics, we briefly review some fundamental concepts and experiments from quantum information theory, such as the Mach-Zehnder interferometer, the Stern-Gerlach experiment, Bell inequalities and entanglement, and the language of quantum circuits.
\end{abstractinenglish}
\begin{keywordsenglish}
Quantum information and gravity; gravitationally induced entanglement; indefinite causal order.
\end{keywordsenglish}

\section{Introduction}

Quantum theory and general relativity are the two fundamental pillars of modern physics. With them, we are able to describe phenomena ranging from the behavior of the smallest particles we know, such as atoms, electrons, and photons, to the dynamics of large cosmic structures, such as stars, galaxies, and black holes.
However, these theories apply to distinct regimes: quantum mechanics describes phenomena in the microscopic world with great precision, while general relativity successfully explains the behavior of the universe on macroscopic scales. To this day, we still do not know how to unify them into a complete and coherent theory.

Quantum Field Theory~\cite{Schwartz_2013} is, so far, our most comprehensive theory, as it unifies quantum physics and special relativity. Attempts to include the gravitational field in this unification by treating it as a quantum system have led to the formulation of several candidates for a theory of quantum gravity, such as string theory~\cite{Abdalla_2005, Mukhi_2011} and loop quantum gravity~\cite{Ashtekar_2021}.
The main focus of these theories is to explain the high-energies regime, such as the formation of black holes and the physics of the early universe, soon after the Big Bang. Therefore, it is difficult to make reasonably testable predictions that would allow us to infer which candidate for a quantum gravity theory would be the most suitable.

However, an even more fundamental question remains open: should gravity really be treated as a quantum system? The truth is that we do not know if quantizing gravity is actually necessary in a world consistent with both theories~\cite{doi:10.1142/S0218271823420245}. An alternative would be to assume that gravity induces modifications in quantum mechanics~\cite{Diosi,Penrose}, sometimes referred to  the ``gravitization'' of quantum theory~\cite{Penrose}. These theories predict interesting phenomena, such as the generation of noise associated with the collapse of the wave function~\cite{Carlesso2022}.

Among so many theories and phenomena that we have not yet been able to verify, is there at least a way to infer which direction would be best to follow in order to unravel the mysteries between quantum theory and gravitation?
In this context, the field of quantum information can play a relevant role, as it is dedicated to characterizing what it means to be quantum and how far this behavior departs from classicality. Thus, the problem of determining whether gravity should be quantized aligns well with the tools developed in this field.

To appreciate how quantum information can contribute to the discussion, it is important to briefly understand what this discipline investigates~\cite{nielsen_chuang}. In general, quantum information studies how information is represented, manipulated, and transmitted in systems governed by the laws of quantum mechanics. Its central ingredients are {\it quantum bits}, or {\it qubits}, which have an algebra much richer than the traditional bits that assume the values $0$ and $1$. A qubit can be in a state of superposition, described by the bases $|0\rangle$ and $|1\rangle$, which allows exploring unprecedented possibilities in information processing. Furthermore, they cannot be copied due to the quantum no-cloning principle~\cite{Wootters1982, Vlado}, which guarantees security for several quantum communication protocols~\cite{Horodecki_2009}. Another fundamental phenomenon is entanglement, which arises from the interaction between two or more qubits and cannot be completely reproduced by classical systems. This resource is one of the main reasons for the efficiency of algorithms in quantum computing and quantum communication protocols~\cite{Horodecki_2009}.

In this article, we will briefly discuss some fundamental results and experiments that allow us to understand why superposition states and entanglement are phenomena that classical physics cannot explain. In particular, we will address the Mach-Zehnder interferometer, which displays the wave-particle behavior of light and matter, and the Stern-Gerlach experiment, which reveals that the spin of elementary particles is discrete. Furthermore, we will discuss Bell inequalities, which allow us to verify and test in a simple and elegant way that quantum phenomena cannot be explained classically. Finally, we introduce the basic elements of the circuit language, which is very useful for describing the evolution of a quantum system and will assist us in later discussions.

These experiments and mathematical tools are well-established. They have various applications in new technologies~\cite{Horodecki_2009} and in the development of the foundations of quantum theory~\cite{RevModPhys.86.419, PLAVALA20231}. Here, however, our focus will be on applying these principles to explore whether the nature of gravity is quantum or not.
Instead of studying a given quantum gravity theory in detail, we aim to infer whether it is indeed necessary to formulate a theory in which gravity is not classical.
To do this, we will show how interferometry, the generation of entanglement, and Bell inequalities can be adapted and generalized to explore properties of the gravitational field, pedagogically exposing the principles of two recent research topics: {\it gravitationally induced entanglement} (GIE)~\cite{Bose,MarlettoVedral} and {\it indefinite causal order} (ICO)~\cite{TeseBruna}.

This material is not meant to be a bibliographic review. It offers an introduction to the interface between quantum information and gravity and points to works that develop the covered topics, interpretations, and research directions in more depth. In the first of these topics (GIE), as the name suggests, the possibility of creating entanglement between two particles from their gravitational interaction is investigated. Proposals for the realization of this experiment are quite concrete, and may be carried out in the near future. But what these experiments will allow us to conclude about the quantum behavior of gravity is a topic of intense debate. The second topic (ICO) seeks to reconcile radical properties of quantum theory and general relativity, focusing on the study of causal structures. One of the studied consequences of this reconciliation is the existence of situations where we cannot attribute a fixed past and future relation between events, and several mathematical tools have been developed to describe these so-called processes with indefinite causal order.

For a better understanding of these topics, we discuss the key elements of quantum information in section~\ref{SecIQ}, which we mentioned above.
In sections~\ref{SecGIE} and \ref{SecICO}, which are independent of each other, we describe how to use and adapt quantum information methods to study gravity. In section~\ref{SecGIE} we introduce the fundamental ideas for the phenomenon of gravitationally induced entanglement, and in section~\ref{SecICO} for indefinite causal order. At the end of each section, we discuss the implications and controversies of each topic. In section~\ref{SecConc}, we present our final considerations.

\section{Principles of Quantum Information} \label{SecIQ}

Here, we review some fundamental elements of quantum information, such as the Mach-Zehnder interferometer, the Stern-Gerlach experiment, entanglement, Bell inequalities, and the circuit language. Our goal is to establish the context and language used in the rest of the text. Although the following is accessible to those that have little familiarity with quantum theory, it does not replace a systematic study of each topic. Readers who are specialized in quantum information can proceed directly to section~\ref{SecGIE} or section~\ref{SecICO}.

\subsection{The Mach-Zehnder Interferometer}\label{SubSecMZ}

Below, we introduce the Mach-Zehnder interferometer and the essential aspects for the following sections. A detailed study of optical interferometry can be found in reference~\cite{OpticsZilio}, while a study focused on the Mach-Zehnder interferometer can be seen in~\cite{Marshman_2016}.

\begin{figure}[t]
  \centering
 \includegraphics[width=1\linewidth]{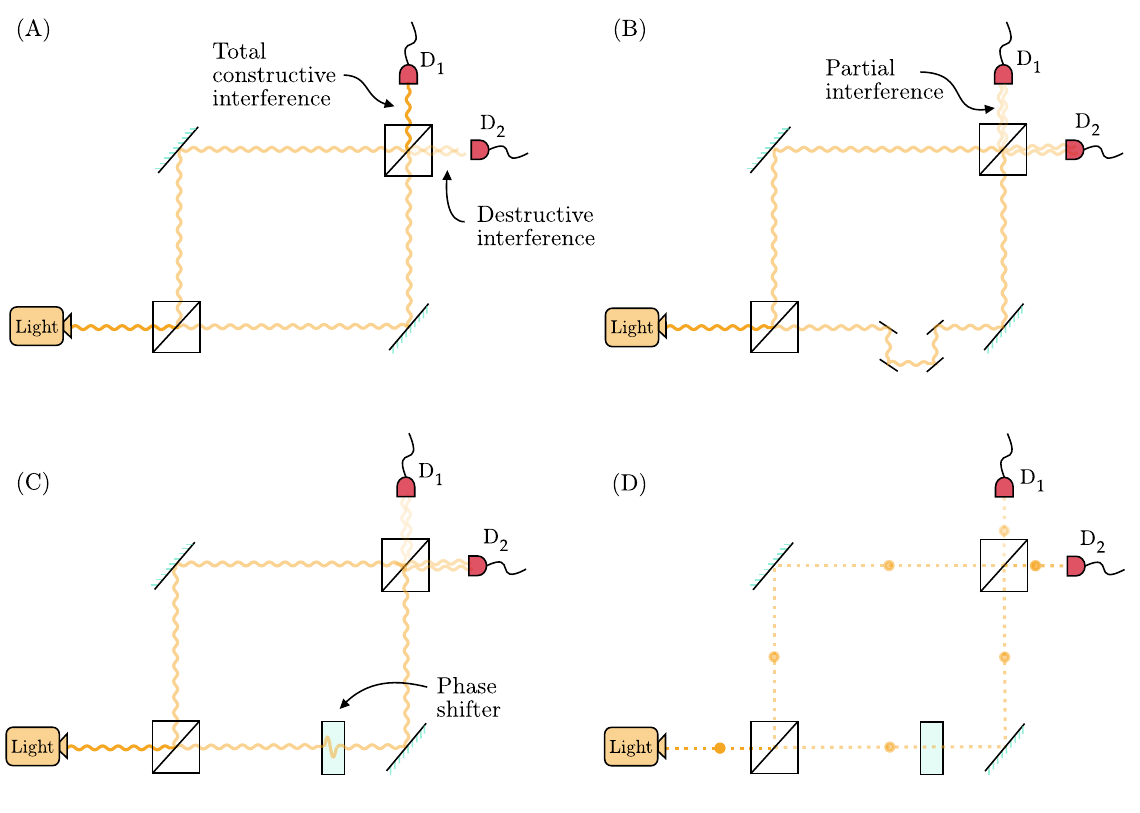}

\vspace{-0.5cm}
   \caption{Basic configuration of a Mach-Zehnder interferometer. Items (A)-(C) represent classical light, that is, a wave, and (D) represents a light particle, that is, a photon. The size difference between the two possible paths for the light defines the detection pattern obtained in D$_1$ and D$_2$. (A) Paths of identical size lead to total constructive interference in D$_1$ and total destructive interference in D$_2$. (B) A path difference is generated by the introduction of a deviation in the lower path, which leads to partial interference in the detectors. (C) The path difference is generated by a phase shifter, which also generates partial interference. (D) Mach-Zehnder interferometer for a particle. The statistics generated after several detections are the same as for a classical wave.
   }\label{fig:MachZehnder}
\end{figure}

The Mach-Zehnder interferometer is a simple optical apparatus consisting of two standard mirrors and two semi-transparent mirrors (beam-splitters) arranged symmetrically on a table. Figure~\ref{fig:MachZehnder} represents the table seen from above. Consider initially figure~\ref{fig:MachZehnder}(A). A light source is placed before the first beam-splitter, located in the lower left part. The light travels through the apparatus to the upper right corner, where there are two detectors that `click' when they receive light. When the light encounters a semi-transparent mirror, there is a 50\% chance that it is transmitted and a 50\% chance of reflection.

If we send a beam of light and make sure that the two paths have the same size, as in figure \ref{fig:MachZehnder}(A), only detector D$_1$ receives light. This is a consequence of the interference effect, a wave property predicted by classical optics and electromagnetism. Due to the phase accumulated in the reflections in each path, the phase difference between the two components of the wave arriving at D$_1$ is zero, generating a total constructive interference. At detector D$_2$, this phase difference is half a wavelength, generating a total destructive interference. That is, D$_2$ does not receive light. This experiment can be seen as a variation of the double-slit experiment, where light waves coming from two sources overlap and undergo interference, generating light and dark fringes on a screen.

If we vary the size of one of the paths as in figure~\ref{fig:MachZehnder}(B), it is possible to make D$_2$ receive a certain amount of light. If the path difference is equal to half a wavelength, the inverse effect occurs: the waves overlap with total constructive interference in the region of D$_2$ and total destructive interference at D$_1$. In practice, varying this length difference little by little is like traversing a pattern of light and dark fringes one piece at a time. It is also possible to do this using a phase shifter, a piece of material that changes the phase of the wave passing through it. This produces an effect analogous to lengthening the path, as illustrated in figure \ref{fig:MachZehnder}(C).

Why does this apparently classical experiment reveal something about quantum theory? The description above becomes completely different when we recognize the basic unit of light, the photon. We know that light has particle behavior thanks to the detection and theoretical explanation of the photoelectric effect in 1905~\cite{EisbergResnick}. This effect corroborates the hypothesis that light only transfers energy in discrete quantities or `quanta'. Consider this same experiment with a single photon, as in figure \ref{fig:MachZehnder}(D). In fact, since the 80s there are sources that emit only one photon at a time~\cite{ASPECT2013315}. In this case, it would be counterintuitive to use wave theory. If a particle has 50\% chance of being reflected and 50\% chance of being transmitted in the beam-splitters, it is expected that each detector clicks roughly half of the time after several trials. However, this does not happen. Even for sources that release one photon at a time, we only observe clicks at the detector D$_1$ 100\% of the time. This is unexpected behavior for a particle and standard classical mechanics fails to describe it. We say that it is evidence of the so-called wave-particle duality of light. But it can be explained by quantum mechanics.

To mathematically describe a quantum experiment, we must consider a Hilbert space $\mathcal{H}$~\cite{Cohen1989}, which is a vector space with an inner product. In the case of the Mach-Zehnder interferometer, where there are two possible paths, we can use a two-dimensional complex vector space $\mathcal{H}=\mathds{C}^2$. Systems represented by this type of space are called {\it qubits}, and their
states are represented by vectors in $\mathds{C}^2$, which can be written in terms of the basis
\begin{equation}\label{01vetores}
   \ket{\text{0}} := \begin{pmatrix} 1\\ 0 \end{pmatrix}, \quad  \ket{\text{1}} := \begin{pmatrix} 0\\ 1 \end{pmatrix}.
\end{equation}
The state $\ket{\text{0}}$ represents the particle in the path that starts at the laser, passes through the upper arm of the interferometer, and ends at detector D$_1$, while $\ket{\text{1}}$ represents the complementary path, which starts below the first beam-splitter, passes through the lower arm, and ends at D$_2$.

With this definition, if $\ket{\psi_\text{f}}$ is the final state of the photon, the probabilities that detector D$_1$ or D$_2$ click are given by the squared of the inner products:
\begin{equation}\label{eqD1D2}
P_{\text{D}_1}=|\bra{0}\ket{\psi_\text{f}}|^2 \quad
P_{\text{D}_2}=|\bra{1}\ket{\psi_\text{f}}|^2.
\end{equation}
If, for example, $\ket{\psi_\text{f}}=\begin{pmatrix} \alpha\\ \beta \end{pmatrix}$, where $\alpha$ and $\beta$ are complex numbers, we have that
$\bra{0}\ket{\psi_\text{f}}=(1,0)\cdot\begin{pmatrix} \alpha\\ \beta \end{pmatrix}=\alpha$ and $\bra{1}\ket{\psi_\text{f}}=(0,1)\cdot\begin{pmatrix} \alpha\\ \beta \end{pmatrix}=\beta$, resulting in $P_{\text{D}_1}=|\alpha|^2$ and $P_{\text{D}_2}=|\beta|^2$.

Thus, assuming that the states of the photon can be represented by the vectors above, we can describe the results of the Mach-Zehnder interferometer with quantum mechanics as follows.

The beam-splitter is represented by the matrix
\begin{equation}
S = \frac{1}{\sqrt{2}}\begin{pmatrix} 1&1\\ 1&-1\end{pmatrix},
\end{equation}
which is an operation on the state vector. Considering that the initial state is $\ket{0}$, it acts as:
\begin{equation}\label{eqBS}
S\ket{0}
=\frac{1}{\sqrt{2}}\begin{pmatrix} 1&1\\ 1&-1\end{pmatrix}\begin{pmatrix}
   1\\0
\end{pmatrix}=\frac{1}{\sqrt{2}}\begin{pmatrix}
   1\\1
\end{pmatrix}= \frac{\ket{\text{0}} + \ket{\text{1}}}{{\sqrt{2}}}.
\end{equation}

This state is a linear combination of states $\ket{0}$ and $\ket{1}$, which also belongs to the Hilbert space. Vectors of this type are called {\it quantum superposition states}.
The photon enters this state after passing through the first beam-splitter, and the probability of detection in each of the paths is 50\%. This would be the statistics obtained if each detector were positioned in one of the arms, before the second beam-splitter. However, when the light reaches the second beam-splitter, the matrix S is applied again, and we obtain the final state:

\begin{equation}
\ket{\psi_\text{f}}= S \left[\frac{\ket{\text{0}} + \ket{\text{1}}}{{\sqrt{2}}}\right] = \begin{pmatrix}
   1\\0
\end{pmatrix} = \ket{\text{0}}, 
\end{equation} which gives us again 100\% chance of detection at D$_1$.

If a phase shifter is added to the interferometer, as in figure~\ref{fig:MachZehnder}(C), we also need to consider the operation that this device performs on path $\ket{1}$. The matrix that represents the phase shifter is given by
\begin{equation}
R = \begin{pmatrix} 1&0\\ 0&e^{i\varphi}\end{pmatrix} ,
\end{equation}
where $\varphi$ is a real number that we call {\it phase}. The value of $\varphi$ depends on the refractive index of the phase shifter and its length in the direction of the optical path. This phase shifter has the following effect on the photon state~(\ref{eqBS}) after passing through the first beam-splitter:
\begin{equation}
R \left[\frac{\ket{0}+\ket{1}}{\sqrt{2}}\right]=
\frac{\ket{0}+e^{i\varphi}\ket{1}}{\sqrt{2}}.
\end{equation}

To obtain the final state of the photon, we must apply the sequence of operations $SRS$ on the initial state $\ket{0}$, which returns
\begin{equation}
\ket{\psi_\text{f}}=SRS\ket{0}
=SR \left[\frac{\ket{\text{0}} + \ket{\text{1}}}{{\sqrt{2}}}\right]
=S \left[\frac{\ket{\text{0}} + e^{i\varphi}\ket{\text{1}}}{{\sqrt{2}}}\right]
= 
\frac{(1+e^{i\varphi})\ket{\text{0}} + (1-e^{i\varphi})\ket{\text{1}}}{{2}}.
\end{equation}
Using equation~(\ref{eqD1D2}), the detection probabilities at D$_1$ and D$_2$ are
\begin{equation}
P_{\text{D}_1}= \frac{1+\cos\varphi}{2},
\quad
P_{\text{D}_2}= \frac{1-\cos\varphi}{2}.
\end{equation}

The Mach-Zehnder interferometer is not restricted only to photons, but any quantum particle can reproduce these results. The Mach-Zehnder has already been carried out with electrons~\cite{IntEletrons, Ji2003}, atoms~\cite{IntAtomosMoleculas}, and even large molecules~\cite{Arndt:1999kyb}. One of the current challenges is to implement variations of this interferometer with increasingly larger systems, such as Bose-Einstein condensates, which contain hundreds of atoms~\cite{InterferometrySpace}.

In section \ref{SecGIE} we will show how we can use this type of interferometry with massive systems to study gravitational interactions between quantum particles.

\subsection{The Stern-Gerlach Experiment}\label{SubSecSG}

The Stern-Gerlach experiment is fundamental to quantum physics, as it demonstrates the quantization of the magnetic angular momentum. In addition, it exemplifies the notion of qubits and the different measurements that can be performed on a quantum system. Here we will stick only to the essential elements for the development of the next sections. An in-depth study of all the particularities of this experiment can be found in reference~\cite{SakuraiNapolitano}, while a more recent approach can be seen in~\cite{Cristhiano_SternGerlach}.

\begin{figure}[h]
   \centering
   \includegraphics[width=1\linewidth]{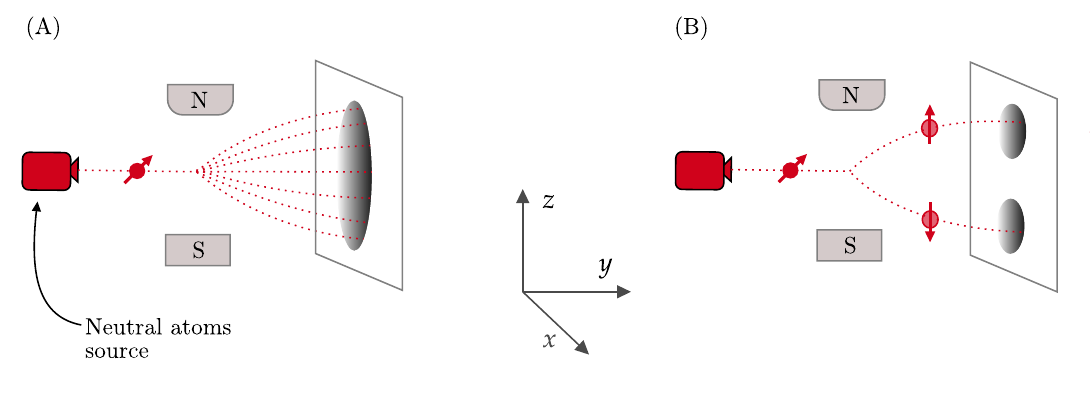}
   \vspace{-1cm}\caption{Stern-Gerlach experiment. (A) Theoretical prediction using classical physics. (B) Experimental observations.}\label{fig:SternGerlach}
\end{figure}

The Stern-Gerlach experiment consists of a source that emits neutral atoms with angular momentum in random directions, a pair of magnets that creates a non-uniform magnetic field in the region where the atoms pass, and a plate that attests the final position of the atoms, as illustrated in figure~\ref{fig:SternGerlach}.

Let us say the atoms are emitted in the $y$ direction, and the magnets are oriented in the $z$ direction. In general, a system with magnetic angular momentum changes the direction of its trajectory in the presence of a non-uniform magnetic field.
If the magnetic angular momentum was a continuous variable as predicted by classical theories, the observations of the Stern-Gerlach experiment would be as illustrated in figure~\ref{fig:SternGerlach}(A), where the atoms collide with the plate in a continuous range along the $z$ direction.

However, the experiment showed that the atoms colide with the plate only in two specific regions\footnote{This result corresponds to experiments using Silver atoms. For atoms with a greater angular momentum, the observations of this experiment would be slightly different.}, one above and one below the emission region, as illustrated in figure~\ref{fig:SternGerlach}(B). This observation indicates that the magnetic angular momentum of elementary particles is discrete.

This is was further discovered to be an intrinsic property of elementary particles called {\it spin}, and should not be confused with a literal rotation of the particle. It is a fundamental quantum angular momentum, which only assumes certain values. In this case, the spin can only assume the values: $+\hbar/2$, which we identify as spin up, and $-\hbar/2$, which we identify as spin down, where $\hbar$ is the reduced Planck's constant. We then generally call this a particle of spin $1/2$.

To mathematically represent the state of a particle with spin $1/2$, we consider a Hilbert space $\mathcal{H}=\mathds{C}^2$, and we represent the spin up and spin down orientations by:
\begin{equation}\label{cimabaixovetores}
   \ket{\uparrow_z} := \begin{pmatrix} 1\\ 0 \end{pmatrix}, \quad  \ket{\downarrow_z} := \begin{pmatrix} 0\\ 1 \end{pmatrix}.
\end{equation}
Note that this is the same representation for qubits that we introduced in equation~(\ref{01vetores}). That is, the mathematics of the states in equations~(\ref{01vetores}) and (\ref{cimabaixovetores}) is the same, and we change nomenclature just to indicate we are dealing with different physical systems.

Note that we can also align the magnets in any direction on the $xz$ plane, and we will obtain an observation similar to figure~\ref{fig:SternGerlach}(B), but aligned with the magnets. We say that the spin is measured in the $z$ direction (or $x$) when the experiment is made with the magnets aligned in that direction. As in the Mach-Zehnder interferometer, some unexpected phenomena occur here, which cannot be described classically.

An important example in this case occurs when we perform sequential measurements. Suppose we measure the spin of an atom in the $z$ direction, and we obtain spin up. If we repeat the measurement in the $z$ direction, we will always obtain the same result of spin up. However, if between two measurements in the $z$ direction, we apply a measurement in the $x$ direction, we will find that the result of the second measurement in the $z$ direction is random, with 50\% chance of being up or down. This behavior indicates that measurements in the $x$ and $z$ directions are incompatible. That is, we cannot have knowledge of the spin of a particle in all directions.

If we name $\ket{\uparrow_x}$ and $\ket{\downarrow_x}$ the states of the particles that follow the upper and lower trajectories in the $x$ direction, the previous statement means, more specifically, that the states $\ket{\uparrow_x}$ and $\ket{\downarrow_x}$ have to be treated necessarily as superposition states in $z$ like:
\begin{equation}
\ket{\uparrow_x}
:=
\frac{\ket{\uparrow_z}+\ket{\downarrow_z}}{\sqrt{2}}
=
\frac{1}{\sqrt{2}}\begin{pmatrix} 1\\ 1 \end{pmatrix}, \quad  \ket{\downarrow_x}
:=
\frac{\ket{\uparrow_z}-\ket{\downarrow_z}}{\sqrt{2}}
=
\frac{1}{\sqrt{2}}\begin{pmatrix} 1\\ -1 \end{pmatrix}.
\end{equation}

The spin measurements in the $z$ and $x$ directions are represented by the matrices
\begin{equation}
\sigma_z=\begin{pmatrix} 1&0\\ 0&-1\end{pmatrix} 
\quad \text{and} \quad
\sigma_x=\begin{pmatrix} 0&1\\ 1&0\end{pmatrix}.
\end{equation}
The expected value of the spin in the direction $i=x,z$ for a particle with spin in the state $\ket{\psi}$ is given by $\bra{\psi}\frac{\hbar}{2}\sigma_i\ket{\psi}$. Note that $\bra{\uparrow_i}\sigma_i\ket{\uparrow_i}=1$ and $\bra{\downarrow_i}\sigma_i\ket{\downarrow_i}=-1$, that is, the spin in the $i$ direction is always up if the particle state is $\ket{\uparrow_i}$ and always down if the state is $\ket{\downarrow_i}$.

In section~\ref{SubSecCHSH}, we use the Stern-Gerlach experiment to exemplify a situation where an agent can choose between two measurements on a quantum system ($\sigma_z$ or $\sigma_x$) with the chance to obtain two possible results for each measurement (spin up or down). Furthermore, in section~\ref{SecGIE}, we discuss how a modification of the Stern-Gerlach experiment can be used to explore gravity. Finally, in subsection~\ref{BellQS}, we will use operations and measurements on the spin of two quantum systems to construct a thought experiment for witnessing indefinite causal order.

\subsection{Quantum Correlations}\label{Sec:Correlacoes}

The experiments described in the previous subsections indicate a difference between probabilities generated by quantum and classical states. It would not be so complicated to accept that systems are neither particles nor waves, but something capable of presenting a fundamentally probabilistic intermediate behavior. However, the statistical strangeness of quantum theory is not limited to interference effects. The interpretation of these predictions becomes even more challenging when more than one physical system is considered.

While quantum theory was born entirely motivated by empirical situations, such as the observation of black-body radiation or the photoelectric effect~\cite{EisbergResnick}, the property of quantum entanglement was first understood theoretically to be later observed. Historically, Einstein, Podolsky, and Rosen were the first to notice that quantum correlations could be highly counter-intuitive. In 1935, these scientists proposed a paradoxical situation that could arise from the quantum description, commonly known as the EPR paradox \cite{EPR}. After much debate, such arguments were mathematically formalized, providing us with a precise understanding of new concepts such as {\it entanglement} and {\it non-locality}.

In the following subsections, we present some of the key components of this story. Detailed introductions to the topics we will discuss can be found in the references~\cite{TeseGlaucia, Scarani}.

\subsubsection{The EPR paradox, hidden variable theories, and Bell inequalities} \label{EPRlambdaBell}

The probabilistic nature of quantum theory raises the question: is the theory complete? For comparison, in thermodynamics we study macroscopic quantities, such as volume and temperature, with no need to determine microscopic variables, such as the position of each molecule that constitute a gas.
The issue is not the nonexistence of these microscopic variables, but rather our lack of knowledge about them. It could be the case that quantum theory follows the same logic, and that there are {\it hidden variables} capable of deterministically explaining quantum phenomena. In this view, quantum theory would be incomplete.

In the EPR paradox, the authors analyze the probabilities obtained when two physically distant systems are measured. They conclude that the measurement on one system would instantly affect the state of the other system, an effect popularly known as ``spooky action at a distance''~\cite{EinsteinBornLetters}.
This gave rise to a lot of debate~\cite{PhysRev.48.696, Schrodinger_1935, BohmAharanov, RevModPhys.29.454, Wigner1961} and was only decisively addressed in 1964, when John S. Bell constructed a simple and elegant mathematical argument that translates the conceptual principles assumed by EPR \cite{Bell}. Bell's argument is as follows: suppose we have two physical systems and two agents, Alice and Bob, who perform measurements on each one of these systems.
Denote by $A_i$ and $B_j$ the measurements that Alice and Bob make on their respective systems, and by $a$ and $b$ the outcomes of these measurements. Let $p(a,b|A_i,B_j)$ be the probability of obtaining outcomes $a$ and $b$ given that measurements $A_i$ and $B_j$ were performed.

Suppose then that nature can be described by a hidden variable theory, and that these variables are represented by $\lambda$.
If the measurements performed by Alice and Bob are indeed independent, the probability of Alice obtaining result $a$ cannot depend on measurement $B_j$ made by Bob, just as the probability of Bob obtaining $b$ cannot depend on $A_i$. Despite that, the probabilities could still have a correlation coming from a common cause, which we will represent by the hidden variable $\lambda$. Therefore, the joint probability should be generally expressed by:
\begin{equation}
p(a,b|A_i,B_j,\lambda)=p(a|A_i,\lambda)p(b|B_j,\lambda).
\end{equation}

If we knew the variable $\lambda$, we could calculate the equation above. But by hypothesis, this is a hidden variable, and we do not have access to it. Let us then consider that there is a probability function $q(\lambda)$ for the variable $\lambda$. Then, we could calculate the probability that Alice and Bob obtain results $a$ and $b$, given that they made measurements $A_i$ and $B_j$, by
\begin{equation}\label{LambdaOculto}
p(a,b|A_i,B_j)=\int d\lambda q(\lambda)p(a|A_i,\lambda)p(b|B_j,\lambda).
\end{equation}
We call correlations that satisfy this condition {\it local correlations}. Otherwise, they are called {\it non-local}.

More systematically, the conditions we discussed above to derive equation~(\ref{LambdaOculto}) are:

{\it (1) Hidden variables:} measurement results are determined by properties that exist a priori, independently of which experiment is realized;

{\it (2) Locality:} the measurement performed by Alice does not affect the measurement performed by Bob, and vice versa;

{\it (3) Free will:} the choice of measurements performed on the system is independent of any other process within the experiment.

Bell showed that probabilities that satisfy conditions {\it (1)--(3)} have certain limitations, which we refer to as {\it Bell inequalities}. With this reasoning, the apparent contradiction presented in the EPR paper comes as a consequence of the implicit assumption that these three hypotheses are jointly true.
Nowadays there is vast literature about this topic, and reviews can be found in~\cite{RevModPhys.86.419, Arash, Belinfante1973HV}.
In the following, we focus on the analysis of a Bell inequality for the case where Alice and Bob have two measurement options which can generate two results.

 

\subsubsection{The simplest Bell inequality} \label{SubSecCHSH}

The CHSH inequality~\cite{CHSH} is the simplest Bell inequality, as it is restricted to the correlation between two physical systems that can be measured in two different ways each, and each measurement can return only two results.
The protocol is as follows: consider two agents, Alice and Bob, and suppose that one qubit is sent to each of them. Independently, they must make a measurement on the state of the received qubit. Finally, each of them must send the information of which measurement was made and the outcome to a third agent, whom we will call Claire.

\begin{figure}[ht]
   \centering
\includegraphics[width=0.25\linewidth]{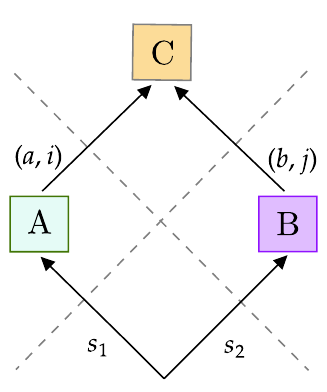}
   \caption{Spacetime diagram for an implementation of the CHSH inequality. Boxes A and B indicate, respectively, the intervals in which Alice and Bob receive systems $s_1$ and $s_2$, perform a measurement, and send the information of which measurement they made and the outcome to Claire. Claire is represented by box C.}
   \label{fig:CHSH}
\end{figure}

If the qubit is the spin of an electron, for example, the agents can choose to measure spin in the $x$ direction or the $y$ direction. The two possible results of each of these measurements are $-1$ (spin down) and $+1$ (spin up).
To ensure that Alice and Bob indeed act independently, we will consider that they are separated by a large enough distance, such that their measurements occur at space-like separated spacetime points, as illustrated in figure~\ref{fig:CHSH}.

We recall that according to relativity theory, two events A and B can have three types of causal relations: A can be in the future of B, A can be in the past of B, or A can be neither in the past nor in the future of B. For this last possibility, we say that A and B have a space-like separation. Thus, two agents with this type of separation cannot communicate.

Consider that Alice can choose between measurements $A_0$ or $A_1$, and Bob between $B_0$ or $B_1$. In the example where our qubit is the spin of an electron, suppose that $A_0$ and $B_0$ are spin measurements in the $x$ direction, and that $A_1$ and $B_1$ are measurements in the $y$ direction. We will denote the result of Alice's measurement by $a$ and Bob's by $b$, which can return the values $+1$ or $-1$.

Given that Alice decides to apply the measurement $A_i$ and Bob the measurement $B_j$, Claire can compute the expected value of the product of the results $a$ and $b$ by
\begin{equation}
\langle
A_iB_j
\rangle
=
\sum_{a,b=-1,1} ab P(a,b|A_i,B_j),
\end{equation}
where $P(a,b|A_i,B_j)$ is the probability that Alice obtains result $a$ and Bob obtains result $b$, given that Alice and Bob implemented measurements $A_i$ and $B_j$.

The CHSH inequality states that if the systems used in the protocol above can be described by a local hidden variable model as in equation~(\ref{LambdaOculto}), or in other words, if it satisfies hypotheses {\it (1)--(3)}, then the statistics of this experiment must satisfy:
\begin{equation}\label{Bell2}
|\langle
A_0B_0
\rangle+\langle
A_0B_1
\rangle+\langle
A_1B_0
\rangle-\langle
A_1B_1
\rangle|
\leq 2.
\end{equation}

Note that such an inequality could be easily violated by disregarding one of the hypotheses {\it (1)--(3)}. For example, if we disregard {\it (2)}, it is possible to violate the inequality above with a classical system, sinces Alice and Bob can communicate and jointly choose how to describe the statistics of their measurements~\cite{PhysRevLett.91.187904}. Requiring the protocol to be done according to the diagram in figure~\ref{fig:CHSH}, this communication cannot happen.

It is remarkable that quantum mechanics can violate the inequality above in situations
where we can guarantee a space-like separation for Alice's and Bob's measurements, and that the choice of these measurements is random. Using quantum systems, it is possible to achieve~\cite{CHSH}:
\begin{equation}\label{Bell2raiz2}
  |\langle
A_0B_0
\rangle+\langle
A_0B_1
\rangle+\langle
A_1B_0
\rangle-\langle
A_1B_1
\rangle|
= 2\sqrt{2}.
\end{equation}
With this, we are forced to conclude that quantum theory cannot satisfy hypotheses {\it (1)-(3)}
simultaneously. Once one is convinced that hypotheses {\it (2)} and {\it (3)} are indeed satisfied,
given the argumentation above, it is concluded that {\it (1)} is false. In other words, we say that quantum theory
is incompatible with a local hidden variable model\footnote{Bohmian mechanics is a theory that explains quantum mechanics predictions using hidden variables, but it is non-local. In this theory, a measurement could influence another one instantaneously, even if they are performed in regions with a space-like separation between them. In this case, hypothesis {\it (2)} is not obeyed.}.

The most powerful part of Bell inequalities is that it is not necessary to consider quantum theory to derive them. For instance, we do not need to define the Hilbert spaces of systems $s_1$ and $s_2$, nor their states or the temporal evolution of their properties. We only consider the probabilities obtained from the joint measurements of Alice and Bob.
We then say that Bell inequalities are {\it theory-independent}. Furthermore, we do not need to know any detail about how the initial state of the physical system was prepared or how Alice and Bob perform their measurements. We only need to know which was the measurement and the outcome. Because of this, we say that Bell inequalities are {\it device-independent}.

\subsubsection{Entanglement}\label{SubSec:Emaranhamento}

To show how quantum theory violates a Bell inequality, we need the concept of entangled states between two physical systems A and B.
Let us consider that the Hilbert spaces associated with these systems are $\mathcal{H}_A$ and $\mathcal{H}_B$, respectively. The joint Hilbert space is represented by the tensor product $\mathcal{H}_A\otimes\mathcal{H}_B$. If, for example, $\mathcal{H}_A=\mathcal{H}_B=\mathds{C}^2$, then $\mathcal{H}_A\otimes\mathcal{H}_B=\mathds{C}^4$. We say that a pure state $\ket{\Psi}_{AB}$ is {\it separable} if it can be written as
\begin{equation}
\ket{\Psi}_{AB}=\ket{\psi}_{A}\otimes\ket{\phi}_{B}.
\end{equation}
In general, we will denote $\ket{\psi}_A\otimes\ket{\phi}_B$ simply by $\ket{\psi}_A\ket{\phi}_B$, or even more concisely, by $\ket{\psi\phi}_{AB}$.
If a pure state of A and B cannot be written in the form above, we say that the system is {\it entangled}.

To violate a Bell inequality in quantum mechanics, it is necessary to use entangled states.
Considering the CHSH scenario, the greatest violation of inequality~(\ref{Bell2}) is given by $2\sqrt{2}$, as in equation~(\ref{Bell2raiz2}). Canonical examples of states that reach this value are known as {\it Bell states}, and are given by
\begin{equation}
\frac{\ket{0}_{A}\ket{0}_{B}\pm\ket{1}_{A}\ket{1}_{B}}{\sqrt{2}},
\quad
\frac{\ket{0}_{A}\ket{1}_{B}\pm\ket{1}_{A}\ket{0}_{B}}{\sqrt{2}}.
\end{equation}

When the physical systems being considered are the spins of two particles, it is common to denote the states above using notation similar to that of equation~(\ref{cimabaixovetores}), and the states above can be rewritten as:
\begin{equation}
\frac{\ket{\uparrow}_{A}\ket{\uparrow}_{B}\pm\ket{\downarrow}_{A}\ket{\downarrow}_{B}}{\sqrt{2}},
\quad
\frac{\ket{\uparrow}_{A}\ket{\downarrow}_{B}\pm\ket{\downarrow}_{A}\ket{\uparrow}_{B}}{\sqrt{2}}.
\end{equation}
For entangled states, we cannot interpret that systems A and B assume independent quantum states. They only admit a complete description when treated jointly.
In effect, separable states can be constructed from operations performed locally on each subsystem, combined with classical communication between the parts,
a set of procedures known by the acronym LOCC (from the English, Local Operations and Classical Communication)~\cite{Horodecki_2009}. It is not possible to create entangled states only from separable states and LOCC.

With these examples, we conclude that quantum theory cannot be reformulated using local hidden variables. We then expect that, if quantum theory indeed provides a good description of the microscopic world, it is possible to perform experiments that violate Bell inequalities.

Nowadays, several experiments that violate these inequalities have already been implemented, and therefore we can conclude that it is not possible to describe them using a local hidden variable theory. This result is so important that it led three scientists who exhaustively tested Bell inequalities in their laboratories to win the Nobel Prize in 2022~\cite{Nobel2022}.

As a final observation, note that there are entangled states that do not violate any Bell inequality. For this, several other mathematical tools have been created to help us verify this property, known as {\it entanglement witnesses}~\cite{TeseGlaucia}.
Unlike Bell inequalities, which are theory-independent, entanglement witnesses are theory-dependent. That is, we can only rely on such quantifiers if we are sure that quantum mechanics describes well the system we intend to characterize as entangled or not. The topic of entanglement witnesses is quite rich, but is outside the scope of this article. A detailed review on the topic can be found in~\cite{TeseGlaucia}.

In sections~\ref{SecGIE} and \ref{SecICO}, we will see how the detection of entanglement and generalizations of Bell's arguments can be adapted for protocols involving quantum systems that interact gravitationally. We will study how these tools can help us explore possibly quantum properties of the gravitational field.

\subsection{Quantum Circuits}\label{Sec:Circuitos}

To describe transformations that are made on quantum systems, we commonly use the representation of {\it circuits}, which is the basic language of quantum computing. This representation is used in the next sections to represent some tasks, and contributes to debates on the physical interpretation of the presented phenomena.

Just as in classical computing, where an electrical circuit is composed of wires and logic gates that transport information from one place to another, the tasks performed on a quantum computer can also be described by wires that represent quantum systems and quantum gates that represent operations performed on these systems. For simplicity, we will consider that our physical systems are qubits. For a more in-depth study on quantum computing, we refer the reader to~\cite{nielsen_chuang, Mermin2003, Mermin2007}.

\begin{figure}[ht]
   \centering
\includegraphics[width=0.85\linewidth]{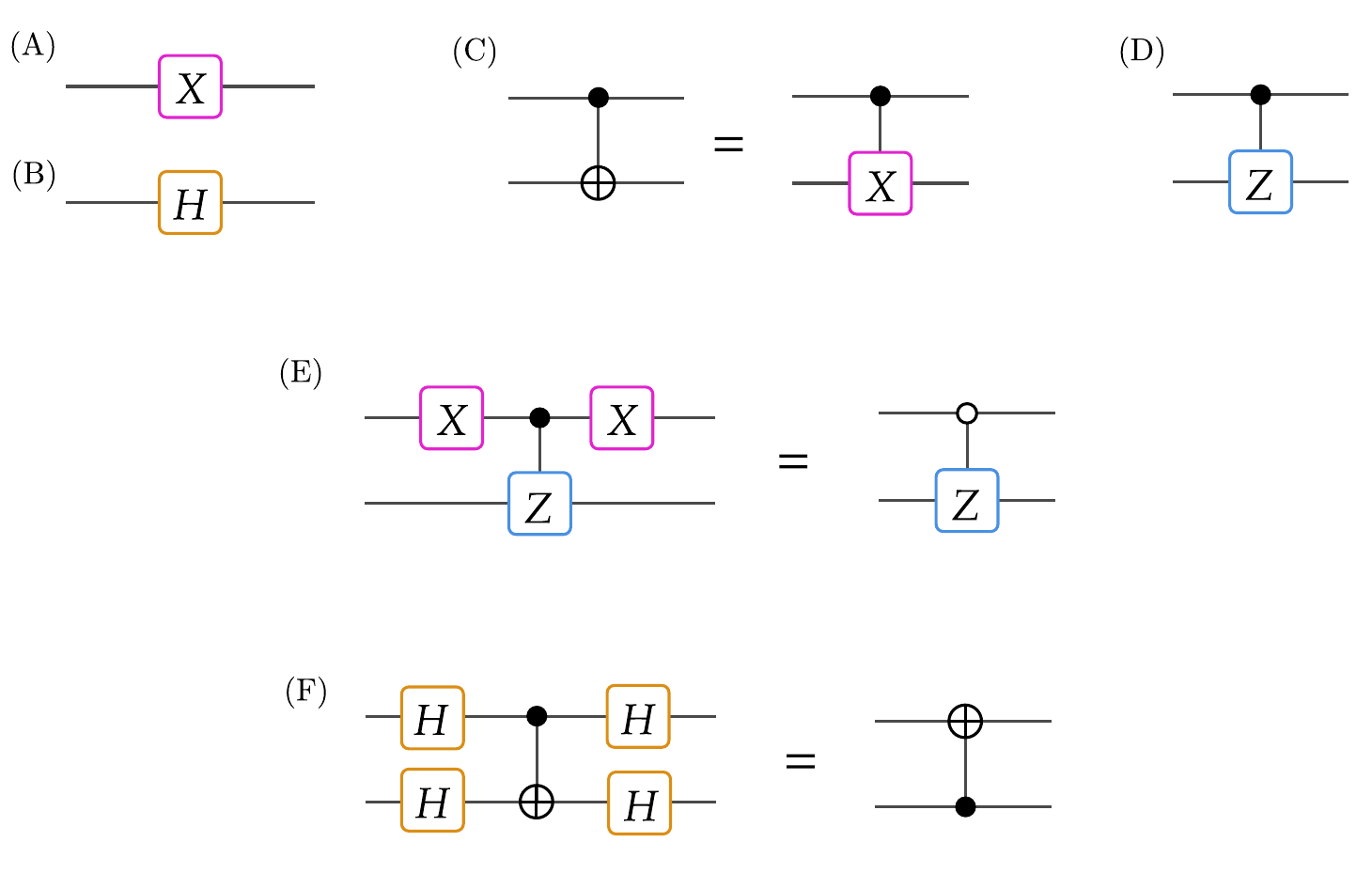}

   \caption{Quantum circuits. (A) Not gate; (B) Hadamard gate; (C) CNot gate; (D) controlled $Z$ operation; (E) composition of Not gates with controlled $Z$ operation; (F) composition of Hadamard gates and a CNot.}
   \label{fig:Circuits1}
\end{figure}

Some simple quantum circuits are represented in figure~\ref{fig:Circuits1}.
By convention, a circuit is read from left to right, and its lines do not necessarily correspond to a physical wire. They can represent the passage of time, or the movement of a particle going from one point to another. In figure~\ref{fig:Circuits1}(A), we have the logic gate known as Not, represented by the letter $X$. If the state of the qubit is $\alpha\ket{0}+\beta\ket{1}$, with $|\alpha|^2+|\beta|^2=1$, its action is to swap coefficients $\alpha$ and $\beta$, returning
\begin{equation}
X(\alpha\ket{0}+\beta\ket{1})= \beta\ket{0}+\alpha\ket{1}.
\end{equation}
Using the vector representation of equation~(\ref{01vetores}), the Not gate must be represented by the matrix
\begin{equation}
X=\begin{pmatrix} 0&1\\ 1&0 \end{pmatrix}.
\end{equation}
Another important logic gate is the Hadamard gate, in figure~\ref{fig:Circuits1}(B), represented by the matrix
\begin{equation}\label{matrizH}
H=\frac{1}{\sqrt{2}} \begin{pmatrix} 1&1\\ 1&-1 \end{pmatrix}.
\end{equation}
It is interpreted as a change of basis of the qubit it acts on. For example:
\begin{equation}
H\ket{0}=\frac{\ket{0}+\ket{1}}{\sqrt{2}}:=\ket{+}, \quad
H\ket{1}=\frac{\ket{0}-\ket{1}}{\sqrt{2}}:=\ket{-} \implies
H(\alpha\ket{0}+\beta\ket{1})=\alpha\ket{+}+\beta\ket{-},
\end{equation}
that is, $H$ applied to a state described by the vector $(\alpha, \beta)$ in the basis $\{\ket{0},\ket{1}\}$, returns the same vector $(\alpha, \beta)$, but in the basis $\{\ket{+},\ket{-}\}$.

A circuit with two qubits is illustrated in figure~\ref{fig:Circuits1}(C). These qubits pass through the Controlled Not gate, or CNot.
We call one qubit the control system and the other the target system. In figure~\ref{fig:Circuits1}(C), the upper wire corresponds to the control qubit, and the lower wire to the target qubit. The operation performed by the CNot gate is: if the control qubit is in state $\ket{0}$, then nothing happens to the target system; if the control qubit is in state $\ket{1}$, then the Not operation is applied to the target qubit. The CNot gate is then represented by the matrix
\begin{equation}\label{matrizCNot}
\text{CNot} = \begin{pmatrix}
  1&0&0&0 \\
  0&1&0&0 \\
  0&0&0&1 \\
  0&0&1&0
     \end{pmatrix}.
\end{equation}

A gate similar to CNot is given in figure~\ref{fig:Circuits1}(D), which represents a controlled $Z$ operation. That is, if the control qubit is in state $\ket{0}$, nothing happens to the target system; if the control qubit is in state $\ket{1}$, then the $Z$ operation is applied to the target. This gate is represented by the matrix
\begin{equation}\label{MatrizCZ}
\text{C}Z=\begin{pmatrix}
  1&0&0&0 \\
  0&1&0&0 \\
  0&0& & \\
  0&0& \multicolumn{2}{c}{\raisebox{.6\normalbaselineskip}[0pt][0pt]{$Z$}}
     \end{pmatrix}
=
     \begin{pmatrix}
1&0&0&0 \\
0&1&0&0 \\
0&0&a&b \\
0&0&c&d
     \end{pmatrix}, \quad
\text{where} \quad
Z=
     \begin{pmatrix}
a&b \\
c&d 
     \end{pmatrix}.
\end{equation}
Note that the CNot gate is a C$Z$ gate with $Z=X$.

Now, if we compose the controlled $Z$ operation with two Not gates, as in figure~\ref{fig:Circuits1}(E), we obtain a controlled $Z$ operation by inverting the vectors $\ket{0}$ and $\ket{1}$. That is, if the control qubit is in state $\ket{1}$, nothing happens to the target system; but if the control qubit is in state $\ket{0}$, operation $Z$  is applied to the target qubit. This protocol is represented by the matrix
\begin{equation}\label{MatrizCZnot}
(X\otimes\mathds{1})\text{C}Z (X\otimes\mathds{1})= \begin{pmatrix}
  &&0&0 \\
  \multicolumn{2}{c}{\raisebox{.6\normalbaselineskip}[0pt][0pt]{$Z$}}&0&0 \\
  0&0&1 &0 \\
  0&0&0&1  
     \end{pmatrix}
=
     \begin{pmatrix}
a&b&0&0 \\
c&d&0&0 \\
0&0&1&0 \\
0&0&0&1  
     \end{pmatrix}.
\end{equation}

Finally, consider the circuit illustrated in figure~\ref{fig:Circuits1}(F).
This circuit is a sequence of Hadamard operations applied separately to each of the qubits, and a CNot between them, in the form:
\begin{equation}\label{HCNotH}
(H\otimes H)\text{CNot}(H\otimes H)
=
\begin{pmatrix}
1&0&0&0 \\
0&0&0&1 \\
0&0&1&0 \\
0&1&0&0
     \end{pmatrix}.
\end{equation}
This operation inverts the role of the control and target qubits: if the target qubit is in state $\ket{0}$, then nothing happens to the control system; if the target qubit is in state $\ket{1}$, then the Not operation is applied to the control qubit.

The circuit representation will be used in subsection~\ref{SUbSecEGIdiscussao} to represent a possible way to understand the interaction between two quantum particles mediated by their gravitational field, in subsection~\ref{SubSecProcICO}, to represent a protocol in which the operations applied to a system happen in an indefinite order, and in subsection~\ref{SubSecDebate} to discuss the notion of causality in this task.

\section{A test of the nature of gravity in quantum regimes} \label{SecGIE}

Quantum information techniques are already used to perform tests of general relativity (classical gravitation) \cite{Hafele1972AroundtheWorldAC, OpticalClocksExp}. Furthermore, there are experiments and experimental proposals that test the effects of classical gravity on quantum systems \cite{Collela,CollelaWerner,GravAharonovBohmEff,Zych_2012,Terno2015,Rivera_Tapia_2020,RelHOM,Roura}. But testing consequences of quantum gravity is, for now, a theoretical investigation. This is due to the scale at which gravitational effects of small systems are important: the Planck scale. Out of this regime, these effects are overshadowed, for example, by electromagnetic forces and the general difficulty of isolating a quantum system from a noisy environment.
Part of the work of physicists is to try to circumvent these difficulties, either by developing new techniques and improving precision on the experimental side, or by formulating protocols on the theoretical side that cause small effects to generate large and measurable consequences.

Recently, the idea that the tension between quantum theory and gravitation can only be tested at the Planck scale has been challenged, at least in its most strict form. More precisely, attention is drawn to the regime where both the gravitational constant (G) and Planck's constant ($\hbar$) are relevant to the experiment, while we can still in principle take the limit in which the speed of light is large: $c\to\infty$. This corresponds to the low kinetic energy or non-radiative regime where one can, for example, consider a superposition of positions of a mass large enough to generate a detectable gravitational field. The theoretical estimate is that, in this situation, quantum effects can be observed even with masses below the Planck mass, on the nanogram scale \cite{quantuminformationmethodsquantumgravity}. In fact, there has been notable experimental progress in recent years both in the control of massive quantum systems \cite{levitacaodenanoparticula,Magrini_2021,Tebbenjohanns_2021}, and in the detection of gravitational effects of progressively smaller masses \cite{Westphal:2020okx} (gold spheres with mass less than 100 mg!). This effort has generated the hope for an experimental test of the nature -- quantum or not -- of the gravity generated by a quantum system in the next decades.

On a theoretical level, these tests have been discussed for some time. For example, in 1957, at the Chapel Hill conference on the role of gravitation in physics \cite{ChapelHillj}, one of the addressed topics was how the gravity of a mass in a superposition of positions would affect another mass in its vicinity, as well as the consistency of the effect with quantum theory. This generated some proposals and realizations of experiments that measure fluctuations of the gravitational force near a mass in a quantum state \cite{bahrami2015gravityquantum,PhysRevLett.47.979,Anastopoulos_2015}, which challenge the main alternative theory to quantization: semiclassical gravity~\cite{Carlip_2008}.

The experiment we are going to describe now is a proposal for detecting an effect called {\it gravitationally induced entanglement} (GIE). It is inspired by the same discussions, and is also placed as a test similar to those mentioned above, but opens the way to richer conclusions. It proposes to test whether gravitational interaction could generate entanglement between two particles and is relatively close to current experimental capabilities. The observation of this effect, in addition to discarding semiclassical gravity theories, could help us infer possibly quantum properties of gravity. Although there is an intense debate about these conclusions, the detection would be a milestone in the study of the nature of the gravitational interaction at quantum scales.

\subsection{Gravitationally induced entanglement}

The experimental proposal to test GIE~\cite{Bose,MarlettoVedral} consists of two massive systems in a superposition of positions, each in a scheme similar to the Mach-Zehnder interferometer in figure \ref{fig:MachZehnder}. There are different ways to prepare this type of configuration and different massive systems could be used. Micro-diamonds and ytterbium crystals are possible candidates for these systems \cite{Bose, Diamonds}. Regardless of the material, the proposals roughly follow the general protocol below.

\begin{figure}[h]
   \centering
\includegraphics[width=0.8\linewidth]{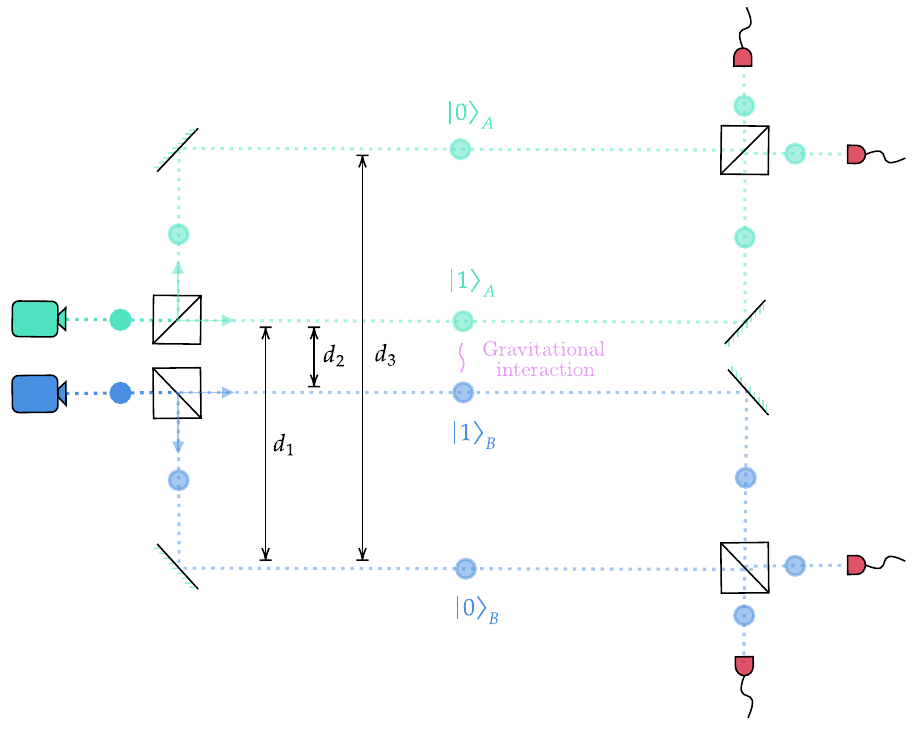}
\caption{Illustration of the experiment to test GIE using two Mach-Zehnder interferometers of massive particles.}
   \label{GIE_MV}
\end{figure}

Consider two matter interferometers \cite{IntAtomosMoleculas}, as in figure \ref{GIE_MV}.
In this setup, the state of massive systems A and B after the first beam-splitter can be written as
\begin{equation}
\frac{\ket{0}_{A}+\ket{1}_{A}}{\sqrt{2}}\otimes  \frac{\ket{0}_{B}+\ket{1}_{B}}{\sqrt{2}}=\frac{\ket{00}_{AB}+\ket{01}_{AB}+\ket{10}_{AB}+\ket{11}_{AB}}{2},
\end{equation}
where $\ket{0}$ and $\ket{1}$ denote the paths in the interferometers. The composite system of two particles is then in a quantum superposition of four possibilities.

If the massive particles interact in some way, the initial state can evolve over time, thus modifying the coefficients, initially $1/2$, associated with each of the vectors above. However, the 50\% probabilities for each particle to be found in one of the two positions halfway through the experiment remains unchanged. In other words, the interaction does not cause the transfer of particles between the arms of the interferometer. Hence, the effect of that interaction on the coefficients can be, at most, a multiplication by a complex number of size 1. A coefficient of this type can always be written as $e^{i\varphi}$ for some value $0\leq\varphi\leq 2\pi$, which we call phase, just as we did in subsection~\ref{SubSecMZ}. This phase can change the results after the recombinations. In this case, it is like one of the interferometers serves as a phase shifter for the other.

The authors also assume that the arms of the interferometers are of the same size and are separated by the same distance $d_3-d_1$. Furthermore, they assume that the distance $d_3$ is much greater than $d_2$ and therefore the only arms close enough to interact gravitationally are those separated by $d_2$ (see Figure \ref{GIE_MV}). It is also assumed that the masses are electrically neutral and that the distance $d_2$ is large enough so that other forms of interaction do not interfere with the experiment \cite{Bose}.

With this, the final state before recombination is restricted to be of the form
\begin{equation}\label{faseBMV}
\frac{\ket{00}_{AB}+\ket{01}_{AB}+\ket{10}_{AB}+e^{i\varphi}\ket{11}_{AB}}{2}, 
\end{equation}
where $\varphi$ is a phase that may depend on the experimental parameters: the mass $m$ of the particles, the distance $d_2$ and the time $t$ itself. It is mathematically possible to show that (\ref{faseBMV}) is generally an entangled state by calculating the entanglement entropy \cite{nielsen_chuang}. This means that it cannot be written as a separable state for any $\varphi$ that is not a multiple of $2\pi$.

For example, we can try treating gravity in the simplest possible way for a quantum scenario. Just add a Newtonian gravitational potential energy term $V(x)= - Gm^2/x$ to the quantum wave equation and treat it as an operator. The equation then would predict that, after a time $t$, the state $\ket{11}$ evolves like \begin{equation}\label{faseNewtoniana}
   \ket{11}_{AB} \mapsto e^{\frac{-iGm^2 t}{{d_{2}} \hbar}}\ket{11}_{AB}.
\end{equation}
Therefore, the phase $\varphi$ in equation (\ref{faseBMV}) should be equal to $-Gm^2t/d_2 \hbar$. This suggests that the observation of an entangled state of two matter particles indirectly indicates quantum behavior of the gravitational interaction.

But the authors in \cite{Bose,MarlettoVedral} use a stronger argument that suggests that any production of entanglement in this experiment would attest that gravity is not classical. An outline of the argument is: suppose we have 3 systems A, B and M. Suppose that interactions can occur between A and M and between M and B, but that A and B do not interact directly. We say that M is a mediator of the interaction between A and B. In this experiment, gravity or the gravitational field plays this role. Assuming, finally, that M is a classical system, then the interaction between A and B is of LOCC type. As we have already mentioned in subsection~\ref{SubSec:Emaranhamento}, if two states are separable and communicate only via LOCC, then they cannot become entangled. Therefore, if M is classical, A and B cannot become entangled. As gravitation does not necessarily obey quantum theory, proofs of more general versions of this statement can be found in the literature, along with the precise definition of ``direct interactions'', ``entangled states'', and what it means to be ``classical'' in those scopes. For instance, a version of the argument was made for generalized probabilistic theories (GPTs) in \cite{FlaminiaGalley}. In section~\ref{SUbSecEGIdiscussao}, we discuss the live debate around these arguments, mentioning the controversies and alternatives raised by different authors.

\begin{figure}[ht]
   \centering
\includegraphics[width=0.6\linewidth]{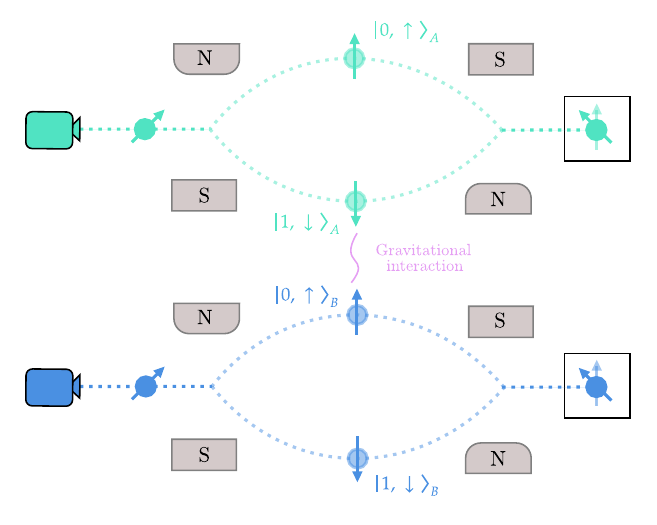}
\caption{Illustration of the experiment to test GIE using an adaptation of two Stern-Gerlach experiments.}
   \label{fig:Bose}
\end{figure}

Quantum information techniques are crucial in the argumentation and experimental considerations. For example, as we mentioned in subsection~\ref{SubSec:Emaranhamento}, we can check whether a quantum state is entangled using an entanglement witness. It can be technically complicated to perform interferometry with objects of mass large enough for this effect to be measurable.
To make the detection of GIE more viable, the authors in \cite{Bose} suggest an adaptation illustrated in figure~\ref{fig:Bose}.

The idea behind this proposal is to use electrically neutral masses that also possess internal spin. The experiment is an adaptation of the Stern-Gerlach experiment introduced in subsection~\ref{SubSecSG}.
Here, two spin 1/2 particles are emitted in the $y$ direction and, for each particle, there is a pair of magnets aligned in the $z$ axis, which change their trajectories according to their spins, as illustrated in figure~\ref{fig:Bose}. Subsequently, for each particle we have another pair of magnets also along $z$, but in the opposite direction. This second pair of magnets deflects the particle's trajectory again, causing the paths to recombine.

It is then assumed that the amount of time during which the trajectories remain separated from each other is much longer than the time necessary to separate them or bring them together. The mechanism in this experiment is similar to the previous proposal: entanglement is generated due to the interaction between the spatially closest trajectories. The difference is that, in this case, the path that each particle went through also depends on its spin.

The state that describes entanglement between the trajectories and spins of the particles before recombination is given by
\begin{equation}
\frac{\ket{0,\uparrow}_{A}\ket{0,\uparrow}_{B}+\ket{0,\uparrow}_{A}\ket{1,\downarrow}_{B}+e^{i\varphi}\ket{1,\downarrow}_{A}\ket{0,\uparrow}_{B}+\ket{1,\downarrow}_{A}\ket{1,\downarrow}_{B}}{2}.
\end{equation}
At the end of the experiment, it is not necessary to perform measurements on the spatial part of the particles. Due to the recombination of the trajectories, the entanglement is all transferred to the state of the spins, which is then expressed as
\begin{equation}
\frac{
\ket{\uparrow\uparrow}_{AB}+ \ket{\uparrow\downarrow}_{AB}+
e^{i\varphi}\ket{\downarrow\uparrow}_{AB}+
\ket{\downarrow\downarrow}_{AB}}{2}. 
\end{equation}
Maintaining the superposition state and verifying entanglement between the spins of the particles is a considerably simpler task.

The protocols discussed here consider entanglement generated in the systems due to their path state. Another proposal for the detection of GIE has also gained attention due to experimental advances in optomechanics \cite{OptoAspelmeyer}. It consists of observing the generation of entanglement between massive harmonic oscillators \cite{Krisnanda_2020, Yant_2023, OscilattorBose_2022}.
\begin{figure}[ht]
   \centering
   \includegraphics[width=0.7\linewidth]{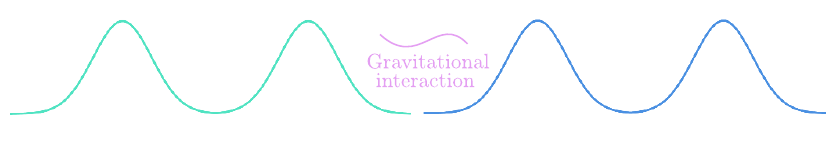}
   \caption{Illustration of the experiment to test GIE using Gaussians in harmonic potentials.}
   \label{fig:enter-label}
\end{figure}

In this type of proposal, the two masses are prepared in a separable state, each in a superposition of two Gaussian states \cite{Ofek_2024}, in harmonic potential traps~\cite{Krisnanda_2020}. They are arranged at a short distance from each other, so they can interact. The mutual gravitational interaction would generate entanglement similarly to the experiments described above, but considering the continuous description of these position states. Proposals like these have paved the way for new ideas that exploit the progress in quantum system control made over the past decades, such as employing a single system to detect its gravitational self-interaction~\cite{GIEcondensado}, among other variants~\cite{quantuminformationmethodsquantumgravity}.

\subsection{Discussions about the interpretation of the experiment} \label{SUbSecEGIdiscussao}

Since the proposals for GIE were made, there has been intense discussion on what we will really learn about gravity if  entanglement is observed in an experimental implementation. There are controversies regarding the original arguments, which are based on a model where gravitational interaction acts as a \textit{mediator}. The debate is built upon the different notions of locality and quantization methods assumed to determine the nature of gravity, in addition to the need for arguments that are theory-independent.

\begin{figure}[ht]
   \centering   
   \includegraphics[width=0.4\linewidth]{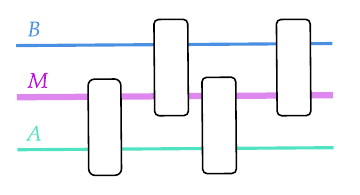}
\caption{Circuit representation of the gravitational interaction between A and B through a mediator M, which illustrates the hypothesis of locality of quantum subsystems.}
   \label{LocSubsys}
\end{figure}

We previously commented on the hypothesis that gravity acts as a mediator M between systems A and B, represented in figure~\ref{LocSubsys}.
In this scenario, A and B do not interact directly, but only through M. This structure characterizes what we call locality of subsystems, an important hypothesis for concluding that M must have a quantum nature~\cite{FlaminiaGalley} in the discussions of the previous section. But does gravity really behave like a mediator?

The idea behind this is to treat the gravitational field as something physical, which propagates signals at a finite speed. For example, if gravity is treated as a Newtonian force, with instantaneous action at a distance, there is no real mediator between A and B \cite{quantuminformationmethodsquantumgravity}. However, since Newtonian gravity is only an approximation and, in general relativity, gravitational effects propagate locally and finitely, the hypothesis of a mediator M can seem reasonable.

But it is important to note that the concept of locality in relativity is different from the concept of locality of subsystems. In relativity, an event can only be influenced by nearby regions, since there is the limit of the speed of light. The link between the two notions is subtle. For example, in quantum field theory, relativistic locality is always maintained, but it is not always possible to represent physics in a region of spacetime as a circuit of the form above \cite{dibiagio2025circuitlocalityrelativisticlocality}. It is even possible to describe what happens in GIE without this circuit structure \cite{MartinezPerche,Perche_2023}. In particular, in \cite{MartinezPerche} they show that it is possible to understand the effect without a mediating system, that is, it would be enough to have a description with ``a classically controlled quantum field'' that does not need a quantum state space to be characterized.

At this point, there is a disagreement between several authors on what can be called ``quantum''. For some, it is enough that the gravitational field can be controlled in a quantum way~\cite{Christodoulou_2023,Fragkos_2022,CHRISTODOULOU201964}. For others, something can only be called quantum if the observed effects cannot be explained without evoking the quantum degrees of freedom of the field \cite{Anastopoulos_2021,MartinezPerche}, a notion derived from usual quantization methods.
In \cite{Fragkos_2022}, the authors explain more about the ambiguity in treating field theories with local or non-local mediators.

Another question is whether we really need an argument that is totally theory-independent. Such an argument would certainly be stronger, but we must remember that there is extensive prior knowledge about the behavior of classical gravity and quantum matter. Inferences based on extrapolations that take these properties into account are nevertheless valuable, despite being valid under more restricted hypotheses.

For example, we have already mentioned that a candidate for the value of the entanglement phase is given by expression (\ref{faseNewtoniana}), derived from a combination of the Newtonian potential with quantum theory. Other types of thought experiments~\cite{Mari_et_al,Belenchia} bring us apparent problems with causality when we perform this treatment, favoring the description of gravity as a quantum field. Other arguments use the linearized gravity theory, and interpret the GIE mediators as gravitons~~\cite{GravitonsBose}, or consider general relativity on microscopic scales to state that the geometry of spacetime itself can be placed in a state of superposition \cite{CHRISTODOULOU201964,Ofek_2024}.

In view of all this discussion, a detection of entanglement in this experiment would be evidence against semiclassical gravity theories where the gravitational field of a mass in a superposition of positions has a definite classical value~\cite{Carlip_2008}, and would also restrict theories that predict gravitational collapse of the wave function \cite{Diosi,Penrose}. If we allow ourselves to make additional hypotheses about the structure of the theory that describes the effect, it could indicate that gravity has a non-classical nature as a field or as spacetime curvature.
Minimally, the detection would show that gravity of particles in superposition states can generate entanglement, a non-trivial information about gravity generated by quantum systems. If (surprisingly) there is no detection of entanglement, this opens up even more space for a new understanding of the nature of gravity.

\section{Causal relations in a non-classical spacetime}\label{SecICO}

In the previous section, we discussed how we can use the generation of entanglement to explore the possibly non-classical nature of the gravitational field. This is one of the ways in which quantum information techniques can be used to explore gravity.
We can also study the consequences of quantum behavior of gravity for the characterization of other gravitational properties, such as the causal structure of spacetime.

In subsection~\ref{Sec:Correlacoes}, we saw how correlations obtained in quantum experiments can be contrasted with those predicted by local hidden variable theories. We can also wonder whether there are ways to distinguish more exotic causal structures from those classically predicted by general relativity. It is interesting to find results like those of Bell inequalities in this context.
Just as quantum theory is not necessary for the formulation of Bell inequalities, we seek to develop theory-independent methods to characterize causal relations.

In the following, we introduce a causal inequality, an expression that can be used to infer whether the causal structure of an experiment is indefinite. After that, we present the canonical example of a task with indefinite causal order, the quantum switch. Then, we discuss Bell's theorem for temporal order, which is the first example of an inequality in this context that has a theoretical violation involving phenomenological hypotheses for quantum gravity. At the end, we review some debates within this research area about both the use of expressions like "causal" or "causality", and the possible interpretations of different protocols.

\subsection{Causal inequality}\label{SubSecDesigualdadeC}

The causal inequality~\cite{Oreshkov} stands out for requiring very few prerequisites to be understood. It characterizes the probabilities involved in a protocol and, under certain hypotheses, its violation would not be consistent with the causal structure of events in a classical spacetime with no closed temporal trajectories~\cite{LoopTemporal, carroll2019spacetime}.
Let us consider a scenario where two agents, Alice and Bob, are in isolated regions, with no possibility of communicating or obtaining information from the external environment, except at two moments: at a first instant, a physical system can enter their region, and, at a later moment, this system leaves. It is highlighted that, in each round of the protocol, the physical system crosses each of Alice's and Bob's regions only once. In this interval, the agents can perform measurements, operations, or encode information in this system.

In addition, Alice and Bob have auxiliary systems that allow them to generate random classical bits. Let us say they are defined by coins in a heads or tails game, and we assign the value $0$ for heads and $1$ for tails. Alice receives a coin and generates a random bit $a$, while Bob receives two coins and generates two random bits $b$ and $b'$.

The protocol proceeds as follows: every time Alice or Bob receive the physical system, they toss their respective coins, and the results define the values of the random bits $a$ or $b$ and $b'$. Each agent, after generating their bit(s), can freely measure or operate on the system, and in particular can encode the value of their bit on it. Before sending the physical system back to the external environment, Alice must try to guess the value of $b$ and Bob the value of $a$, each of them taking note of their guesses. We will denote Alice's guess by $x$ and Bob's guess by $y$.

The goal of the protocol is for Alice and Bob to make the following probability of success as high as possible:
\begin{equation}
p_{\text{suc}}:=\frac{1}{2}
  \left[
   P(x=b|b'=0)
   +
   P(y=a|b'=1)
  \right].
\end{equation}
That is, when $b' = 0$, the agents win the round if Alice correctly guesses the value of $b$. When $b' = 1$, they win if Bob correctly guesses the value of $a$.

Let us assume that, in each round, Alice's and Bob's actions happen in a definite order. For example, suppose Alice acts in the future of Bob. If the two agents are trying to win the game, she receives the physical system that Bob sends already encoded with bit $b$. In this case, Alice can access this information and guess the value of $b$ with $100\%$ certainty, that is: $P(x=b)=1$, and therefore\footnote{As the probabilities of the values acquired by $x$ and $a$ are independent, we have $P(x=b,b'=0)=P(x=b)P(b'=0)$, and by Bayes' rule we obtain that $P(x=b|b'=0)=P(x=b,b'=0)/P(b'=0)=P(x=b)$.} $P(x=b|b'=0)=1$.
However, in this situation, Bob is not in the future of the moment when Alice sends her physical system, and the most he can do is randomly choose the value of $y$. Therefore, there is a 50\% chance of him guessing the value of $a$ correctly, that is: $P(y=a)=1/2$, so $P(y=a|b'=1)=1/2$. Analogous reasoning occurs by reversing the order of events between Alice and Bob.

Acting optimally, Alice and Bob can win this game with a probability of $p_{suc}=\frac{1}{2}(1+\frac{1}{2})$, that is, 75\% of the time. Thus, in a scenario with definite causal order, the probability of Alice and Bob winning the task is bounded by:
\begin{equation}\label{p34}
p_{\text{suc}}\leq \frac{3}{4}.
\end{equation}
Note that in the paragraph above we offered only a general idea that this inequality is valid. A formal proof can be verified in references~\cite{TeseBruna, Oreshkov}.

Equation~(\ref{p34}) is what we call a causal inequality. In the same way as the CHSH is a type of Bell inequality, this is just one type of causal inequality. It was the first formulation of such an idea and was followed by other types, like in~\cite{Branciard_2016, Oreshkov_2016, PinzaniQS, LugtQS}. Note, also, that the causal inequality is theory-independent and device-independent, just like Bell inequalities.

\subsection{Processes with indefinite causal order} \label{SubSecProcICO}

Causal inequalities define certain limitations for processes with definite causal order. But what is the motivation for thinking about a scenario where the order between events is not definite? Naively, one of the motivations is the extrapolation of general relativity principles in specific quantum regimes. For example, if the causal structure of spacetime is defined by curvature, which is determined by the state of matter, a possible expectation for the behavior of the causal structure around a massive quantum system in a superposition state is that it is not definite. Another possibility is that there is some type of collapse, or a classical description of gravity even in quantum regimes. Therefore, the distinction between definite and indefinite causal structures becomes conceptually important.

More generally, it can be argued that general relativity and quantum theory have properties that are revolutionary when compared to previous physical theories. General relativity establishes dynamic causal relatios, being necessary to solve Einstein's equations to determine whether one event is in the past or future of the other. Meanwhile, quantum theory is intrinsically probabilistic in its predictions. Indefinite order arises naturally in formalisms that attempt to reconcile these properties, that is, that present a causal structure that is not fixed a priori and is probabilistic~\cite{Hardy}.

As examples, we have the causaloid formalisms~\cite{Hardy}, supermaps~\cite{Chiribella}, and process matrices~\cite{Oreshkov}.
The idea behind them is to abandon the need to a priori define a global spacetime on top of which phenomena occur, and deal only with regions where we can assume that physics is governed by previously known laws. These regions, also called {\it closed laboratories}, are isolated from the world except for two instants. The first instant is the moment when a physical system enters this region, and the second, when it leaves, as formulated in the causal inequality protocol.

By renouncing the notion of a global spacetime, the traditional definition of what an event would be in relativity theory is also lost, making it necessary to adopt a different notion. According to the process matrix formalism, an {\it event} is given by a potential operation that an agent in a closed laboratory can perform on a quantum system that passes through this laboratory only once.
Note that this new definition may not be able to encompass a complete theory that unifies quantum theory and gravitation, as it is restricted to regions where it is possible to define a closed laboratory. However, we recall that our primary objective is not to search for a complete theory of quantum gravity, but rather to look for scenarios in which we can use quantum information techniques to learn more about gravity. Therefore, this definition of event is sufficient for now.

\begin{figure}[h]
   \centering
   \hspace{-1cm}\includegraphics[width=0.8\linewidth]{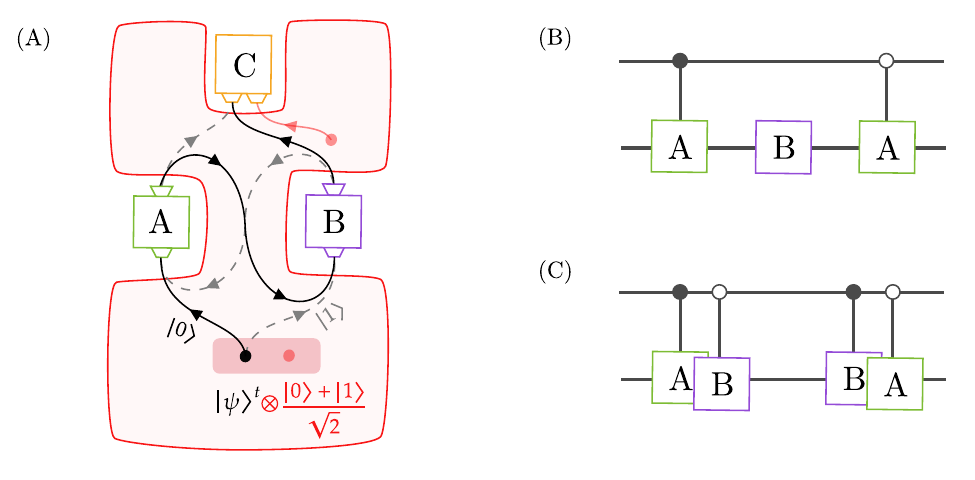}

  \caption{(A) Diagrammatic representation of the quantum switch process. The labels A and B in the boxes represent an event in each of Alice's and Bob's laboratories. (B) and (C) are possible effective representations of the quantum switch in the circuit language.}
   \label{FigQS}
\end{figure}

Several types of processes with indefinite causal order have been mathematically characterized, in particular, processes that violate causal inequalities~\cite{Oreshkov, Lugano}. However, we know physical interpretations only for a certain process and its variations: the {\it quantum switch}. This process is composed of a {\it target system}, a {\it control system}, and three agents, called Alice, Bob, and Claire. Alice and Bob have access only to the target system, on which they can perform generalized operations, while Claire has access to the target and the control system, and can perform measurements on both. Claire is in the future of Alice and Bob, but the causal relation between Alice and Bob is not previously established.

The quantum switch does not violate the causal inequality proposed in~\cite{Oreshkov}, nor any inequality that is formulated in a similar way~\cite{Araujo2015, Wechs21, PurvesShort}. To characterize the quantum switch, it is necessary to use other types of mathematical tools, such as {\it causal witnesses}, which are the analogue of entanglement witnesses for causal order.
However, a simple variation of the quantum switch violates other types of inequality, such as those proposed in~\cite{PinzaniQS,LugtQS}, or in~\cite{tbell}, which we will introduce in subsection~\ref{BellQS}.

For simplicity, let us consider that the target system and the control system are both qubits. In their laboratories, the agents Alice and Bob each apply an operation to the target qubit, which we will denote by $\mathcal{A}$ and $\mathcal{B}$ respectively. The order in which these operations happen is entangled with the control qubit: if the state of this qubit is $\ket{0}_c$, then $\mathcal{A}$ is applied before $\mathcal{B}$, and if its state is $\ket{1}_c$, $\mathcal{B}$ is applied before $\mathcal{A}$. We then have that the following map defines the switch operation:
\begin{equation}\label{QSunitarias}
S_{\mathcal{A},\mathcal{B}}=\ketbra{0}_c\otimes \mathcal{B}\circ\mathcal{A}+\ketbra{1}_c\otimes\mathcal{A}\circ\mathcal{B}.
\end{equation}
If the initial state of the control qubit is $(\ket{0}_c+\ket{1}_c)/\sqrt{2}$ and the target is $\ket{\psi}_a$, the final state of the composite system is:
\begin{equation} \label{QSfinalSt}
S_{\mathcal{A},\mathcal{B}} \left(\frac{\ket{0}_c+\ket{1}_c}{\sqrt{2}}\otimes \ket{\psi}_a\right)
=
\frac{\ket{0}_c\otimes
\mathcal{B}\circ\mathcal{A}\ket{\psi}_a+\ket{1}_c\otimes\mathcal{A}\circ\mathcal{B}\ket{\psi}_a}{\sqrt{2}}.
\end{equation}
By performing a post-selection of the control system in the basis $\{\ket{0}_c\pm\ket{1}_c\}$, the target system is found in one of the states
\begin{equation}
\frac{\mathcal{B}\circ\mathcal{A}\ket{\psi}_a\pm\mathcal{A}\circ\mathcal{B}\ket{\psi}_a}{\sqrt{2}}.
\end{equation}

A diagrammatic scheme for the quantum switch is shown in figure~\ref{FigQS}(A). In this figure, boxes A and B need to be understood as events, and therefore must appear only once in this representation. However, this process can also be obtained with the circuit structures illustrated in figures~\ref{FigQS}(B) and (C), which represent circuits with controlled operations. We recall that controlled operations were introduced in equations~(\ref{MatrizCZ}) and (\ref{MatrizCZnot}) of section~\ref{Sec:Circuitos}. In this case, at least one of the boxes
appears twice in the same representation, which for many authors cannot be treated as the realization of a single
event~\cite{PaunkVoji}.

\begin{figure}[h]
   \centering
   \hspace{-1cm}\includegraphics[width=0.8\linewidth]{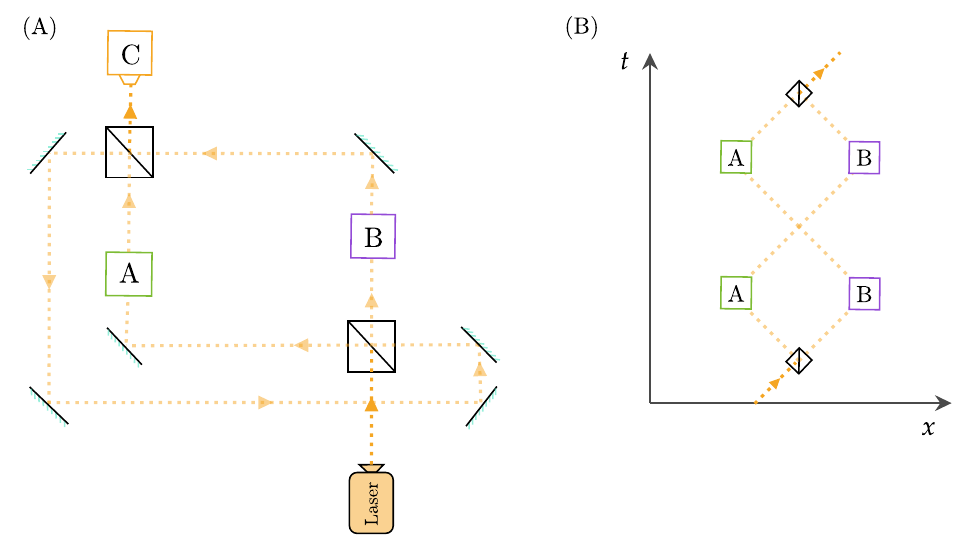}
   \caption{(A) Implementation scheme and (B) spacetime diagram of the optical quantum switch from reference \cite{Goswami}.}
   \label{FigOptQS}
\end{figure}

Note that the entire construction made so far was motivated by the study of causal structures to explore spacetime. However, quantum systems can be delocalized, and this ingredient alone is enough to make the realization of the quantum switch~(\ref{QSunitarias}) possible in a classical spacetime~\cite{ReviewQSopt, Rindler, QSonEarth},
according to the circuit representations~\ref{FigQS}(B) and (C).
In particular, optical experiments that reproduce the effect have been built~\cite{ReviewQSopt} and the map (\ref{QSunitarias}) has been experimentally characterized. As a brief introduction to the optical quantum switch, we summarize here a simplified version of the experiment reported in reference~\cite{Goswami} and illustrated in figure~\ref{FigOptQS}(A).

The control qubit and the target system are both encoded in the same photon, the former in its polarization and the latter in its transverse spatial modes. The photon passes through the optical path illustrated in figure~\ref{FigOptQS}(A). Initially, the photon encounters a polarization beam-splitter. If its polarization is vertical, it is reflected; if it is horizontal, it is transmitted. In the case in which the polarization is vertical, it first encounters box A, which applies an operation $\mathcal{A}$. After being reflected by some mirrors, the photon encounters box B which applies an operation $\mathcal{B}$. If the polarization is horizontal, the photon first encounters box B and then box A, so that this time operation $\mathcal{B}$ is performed before operation $\mathcal{A}$. At the end of the optical path, the photon encounters another polarization beam-splitter that recombines the two paths. The polarization states and the transverse spatial modes of the photon are then measured in box C.
Several rounds of this experiment are performed varying the photon states and also operations $\mathcal{A}$ and $\mathcal{B}$, so that the process as a whole can be characterized.

It is worth noting here that operations $\mathcal{A}$ and $\mathcal{B}$, despite being performed by the same agents A and B on the two paths of the photon, were performed in different spacetime events, as illustrated in figure~\ref{FigOptQS}(B). This reasoning, along with the circuit representation in figures~\ref{FigQS}(B) and (C), marked the beginning of a long debate about the optical quantum switch being a task with genuine indefinite causal order or just a simulation, which we will further detail in section~\ref{SubSecDebate}. Before that, we will present a thought experiment to understand how non-classical gravity, under some basic hypotheses, could generate a spacetime with an exotic causal structure, enabling the realization of a quantum switch.

\subsection{The gravitational quantum switch}\label{SubSecQSG}

The construction of the causal inequality and the definition of the quantum switch were both made with the following definition of event: the application of an operation on a physical system made inside a closed laboratory. However, we have not yet described how this could have a direct connection with a non-classical spacetime, as suggested by the motivations mentioned in section~\ref{SubSecProcICO}. For this, we will describe a thought experiment in which the realization of a quantum switch, under certain phenomenological assumptions, is induced by the gravity generated by a massive object in a superposition of positions.

To introduce the gravitational quantum switch, the authors of~\cite{tbell} consider that each agent has a clock capable of measuring their proper time, and an event inside a laboratory must correspond to a specific instant of time according to their clocks. Note that this definition may or may not be compatible with the definition of event in the process matrix formalism, which we discussed in the previous section, and that is further discussed in section~\ref{SubSecDebate}.

According to general relativity, the presence of massive systems can alter the rate at which time passes, an effect known as gravitational time dilation. For example, suppose Alice is close to a spherical massive body and another agent, Bob, is very far from both Alice and the mass. The relation between the times $\tau$ and $t$ elapsed according to Alice's and Bob's clocks is given by~\cite{carroll2019spacetime}
\begin{equation}
\tau=g_{00}(R) \ t,
\end{equation}
where
\begin{equation}
g_{00}(R)=\sqrt{1-\frac{2GM}{c^2R} \ }
\end{equation}
is one of the components of the spacetime metric, where $G$ is the gravitational constant, $M$ is the mass of the spherical body, $c$ is the speed of light, and $R$ is Alice's distance to the center of the spherical body.

\begin{figure}[ht]
   \centering
   \includegraphics[width=0.95\linewidth]{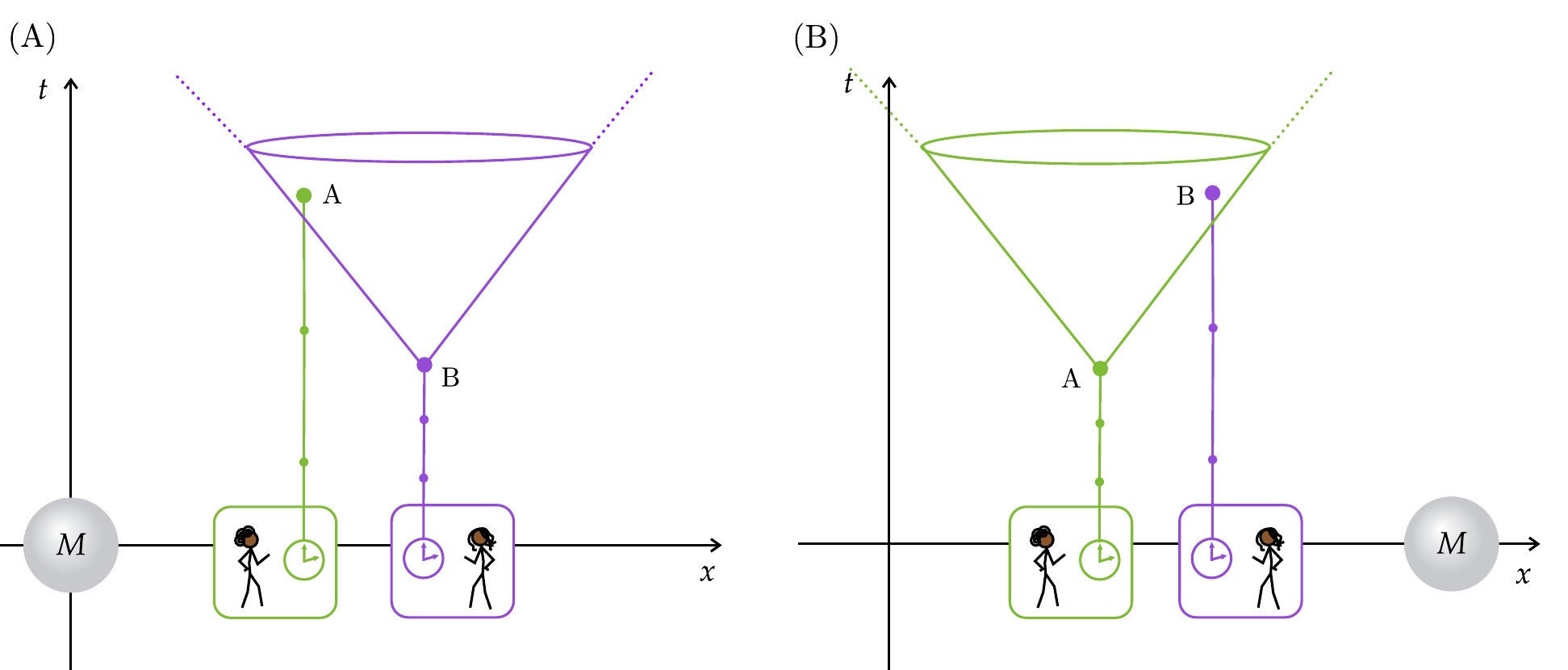}

\vspace{-0.3cm}
   \caption{Illustration of two situations where a massive body is closer to one agent than to the other. (A) Alice is closer to the massive body than Bob, and therefore her proper time runs slower than Bob's. (B) Inverse situation where Bob is closer to the mass than Alice.}
   \label{ABMtime}
\end{figure}

Let us now consider that Alice and Bob are at a distance $R_\text{A}$ and $R_\text{B}$ from the center of the spherical body of mass $M$. We illustrate this situation in figure~\ref{ABMtime}, with $R_\text{A}<R_\text{B}$ in~\ref{ABMtime}(A), and $R_\text{A}>R_\text{B}$ in~\ref{ABMtime}(B). Suppose Alice and Bob synchronize their clocks at a given initial instant and each of them sends a light signal to the other at instant $\tau^*$, according to their clocks. Is it possible that one of them sends their signal after having already received a signal from the other? In reference~\cite{tbell}, it is shown that, if Bob is closer to the mass than Alice, he would receive her signal before sending his own if
\begin{equation}
\tau^*\geq T_c \frac{\sqrt{-g_{00}(R_\text{B})}}{1-\sqrt{\frac{g_{00}(R_\text{B})}{g_{00}(R_\text{A})}}},
\end{equation}
where $T_c$ is the time it takes for a light signal to travel from Alice to Bob. If Alice is closer to the mass, the equation above is valid by swapping indices A and B.
In this way, we exemplify how the presence of a massive body can influence the causal structure between Alice and Bob: if the mass is on the left, event A is in the causal future of B; otherwise, it is in its causal past.

We will show how to use time dilation as a resource to induce the realization of a quantum switch. Consider that the massive body is actually in a superposition state of  the configurations (A) and (B) in figure~\ref{ABMtime}. In principle, to consider a quantum state of a massive body and its gravitational field, we would need a quantum gravity theory. But our intention is to work under very general hypotheses to understand the basic consequences of possibly quantum behavior of gravity.

So, instead of using a specific theory, we only consider the following hypotheses:

{\it (I)} Distinct macroscopic states correspond to orthonormal quantum states;

{\it (II)} Time dilation in the classical limit reduces to what is predicted by general relativity;

{\it (III)} The superposition principle is valid, regardless of the mass of the systems involved.

\begin{figure}[ht]
   \centering
   \includegraphics[width=0.5\linewidth]{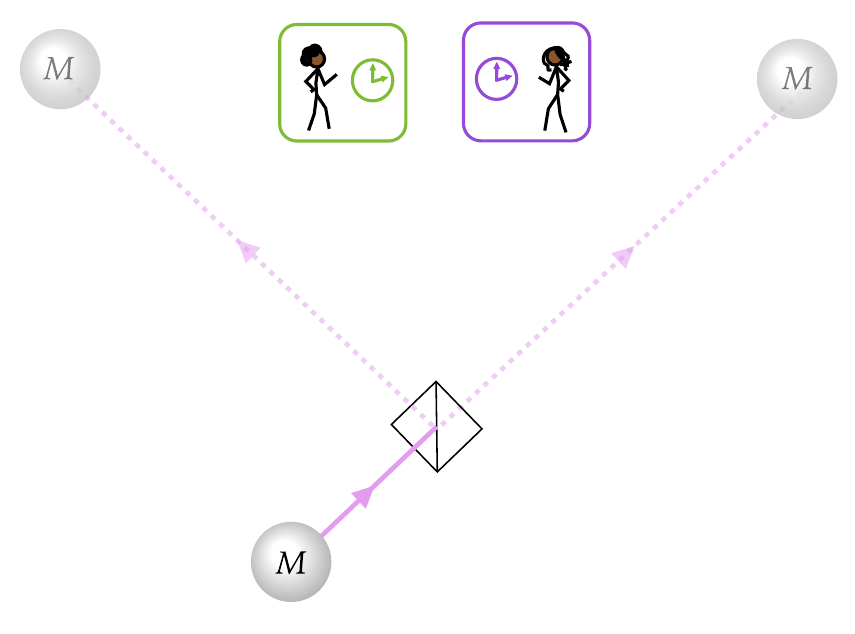}
   \caption{Illustration for the scenario where a massive body is in a superposition of positions. In one case the mass is closer to Alice and, in the other, it is closer to Bob.}
   \label{ABMsuperp}
\end{figure}

With these hypotheses, we can associate an orthonormal quantum state to each of the configurations in figure~\ref{ABMtime}. We will denote by $\ket{K_{A\prec B}}_M$ the state in which the massive body is closer to Bob (figure~\ref{ABMtime}(B)), in which case the signal sent by Alice reaches Bob before event B; and we will denote by $\ket{K_{B\prec A}}_M$ the state in which the massive body is closer to Alice (figure~\ref{ABMtime}(A)). Consider, then, the situation where the initial state of these configurations is in a superposition state, as illustrated in figure~\ref{ABMsuperp}, and described by the expression
\begin{equation}\label{MassSuperp}
\frac{\ket{K_{\text{A}\prec \text{B}}}_M+\ket{K_{\text{B}\prec \text{A}}}_M}{\sqrt{2}}.
\end{equation}

To perform the quantum switch, we will consider the following. Suppose that the event corresponding to $\tau^*$ of Alice is in the past of the event corresponding to $\tau^*$ of Bob. Thus, Alice can perform an operation $\mathcal{A}$ on a physical system with initial state $\ket{\psi}$, and send it to Bob, who performs an operation $\mathcal{B}$ on the system. The final state of this system is $\mathcal{B}\circ\mathcal{A}\ket{\psi}.$ But if Bob's event is in the past of Alice's, an analogous operation can be performed, and the final state of the physical system is given by $\mathcal{A}\circ\mathcal{B}\ket{\psi}$.

Once we consider that the massive body is in the initial state~(\ref{MassSuperp}), we have that the joint final state of the massive body and the physical system shared by Alice and Bob is given by the expression
\begin{equation}
\frac{\ket{K_{\text{A}\prec \text{B}}}_M\otimes \mathcal{B}\circ\mathcal{A}\ket{\psi}_a+\ket{K_{\text{B}\prec \text{A}}}_M\otimes \mathcal{A}\circ\mathcal{B}\ket{\psi}_a}{\sqrt{2}},
\end{equation}
whose form is analogous to equation~(\ref{QSfinalSt}). Then, we can say that this is a protocol for performing a quantum switch using the gravitational field as a control system. Therefore, this protocol is called gravitational quantum switch.

As we mentioned above, the quantum switch does not violate any causal inequality. Therefore, other types of inequalities were formulated to characterize protocols involving the quantum switch \cite{tbell, PinzaniQS, LugtQS}.
We will show in the next section a Bell-type theorem for indefinite order, formulated in~\cite{tbell}, referred to as {\it Bell's theorem for temporal order}.

\subsection{A Bell-type inequality for temporal order} \label{BellQS}

To introduce the gravitational quantum switch in the previous section, it was necessary to make considerations {\it (I)--(III)} about the spacetime that would be generated by a massive object in a superposition of positions. But would it be possible to test if the order is really indefinite without needing to consider such hypotheses? For this, the authors of reference~\cite{tbell} introduced {\it Bell's theorem for temporal order}.

\begin{figure}[t]
   \centering
   \includegraphics[width=0.97\linewidth]{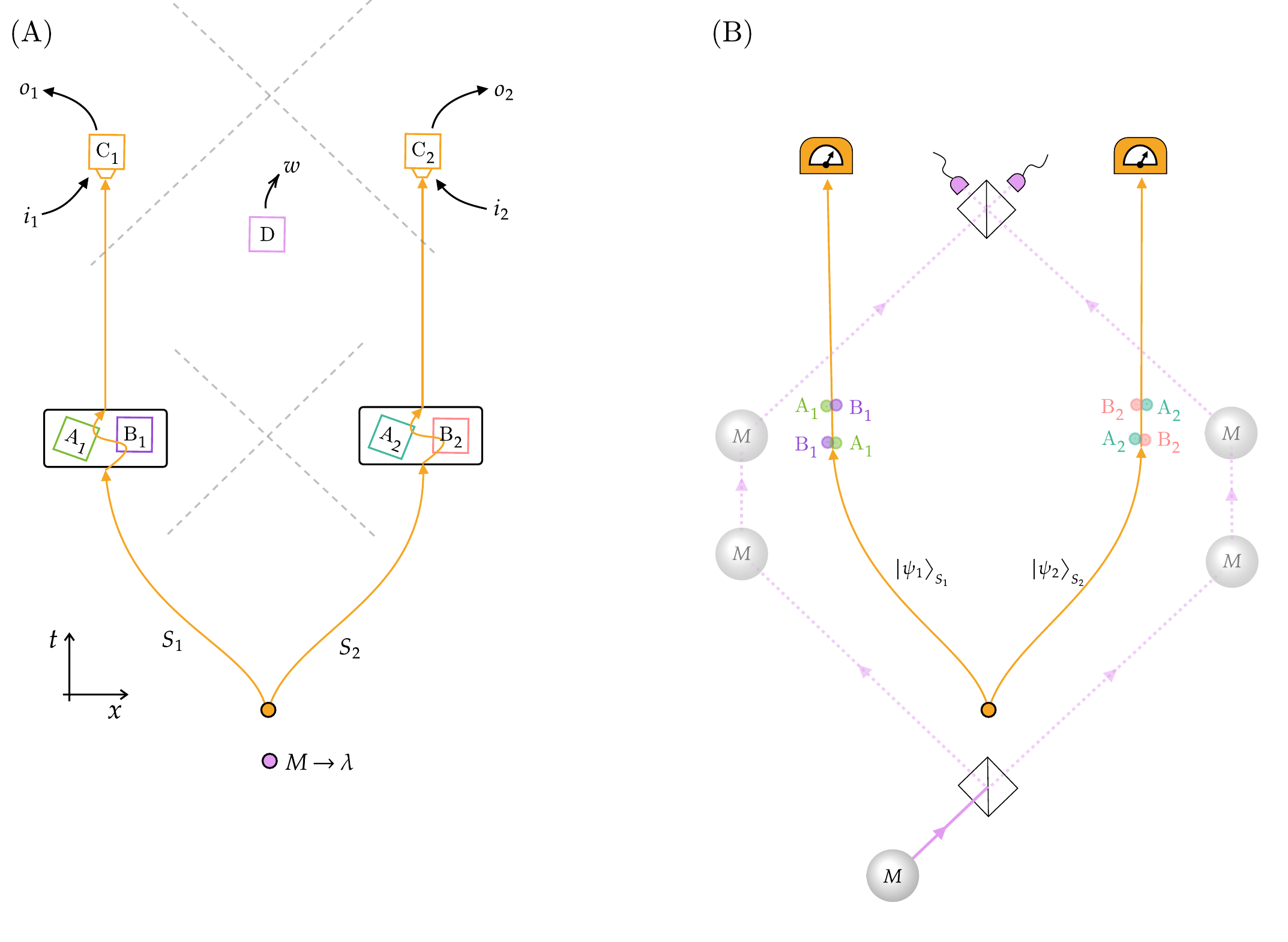}\vspace{-0.7cm}
   \caption{(A) Diagram for the scenario used in Bell's theorem for temporal order; (B) illustration of a thought experiment that violates Bell's theorem for temporal order.}
   \label{fig.BellITO}
\end{figure}

Consider the scenario illustrated in figure~\ref{fig.BellITO}(A). We have a system composed of two parts $S_1$ and $S_2$, a third system $M$, and four agents: Alice$_1$, Bob$_1$, Alice$_2$ and Bob$_2$. The pair of agents
Alice$_n$ and Bob$_n$ can each operate only once on system $S_n$. Let us say that, in event $A_n$, Alice$_n$ applies operation $\mathcal{A}_n$ to $S_n$
and that, in event $B_n$, Bob$_n$ applies $\mathcal{B}_n$ to $S_n$. Finally, in event $C_n$, system $S_n$ undergoes measurement $i_n$ returning the outcome $o_n$. The system $M$ is measured in event $D$, returning outcome $w$.

Bell's theorem for temporal order states that, in an experiment constructed according to the scenario described above and that meets the following hypotheses, the joint state of systems $S_1$ and $S_2$ cannot violate a Bell inequality, regardless of the value attributed to $w$:

{\it (i) Separable initial states:} The initial states of $S_1$, $S_2$, $M$ share no correlations with each other;

{\it (ii) Temporally localized operations:} Agents Alice$_n$ and Bob$_n$ perform their respective operations on system $S_n$ at one specific instant of proper time as measured by their clocks;

{\it (iii) Classical order:} The events in which the operations are performed are classically ordered, that is, for each pair of events $A_n$ and $B_n$, there exists a space-like surface and a classical variable $\lambda$ defined on it that determines the order between events $A_n$ and $B_n$;

{\it (iv) Locality:} Operations $\mathcal{A}_1$ and $\mathcal{B}_1$ do not affect $\mathcal{A}_2$ and $\mathcal{B}_2$, and vice versa. Analogously, measurements $C_1$, $C_2$, and $D$ do not affect each other;

{\it (v) Free will:} The choice of each measurement to be performed in events $C_1$ and $C_2$ is independent of any other process within the experiment.

Note that this theorem is directly inspired by Bell's theorem. Recalling hypotheses {\it (1)}-{\it (3)} introduced in section~\ref{EPRlambdaBell}, it is possible to see that {\it (iii)} introduces the notion of hidden variables for causal order, just as {\it (1)} introduces hidden variables in general, {\it (iv)} gives us a notion of locality just as in {\it (2)}, and {\it (v)} of free will, just as in {\it (3)}. Hypotheses {\it (i)} and {\it (ii)} do not have a direct analogy with the hypotheses for Bell's theorem. Note that hypothesis {\it (ii)} is what connects the order between operations $\mathcal{A}_n$ and $\mathcal{B}_n$ to the causal structure of spacetime.

To violate this inequality, the authors of reference~\cite{tbell} consider two copies of the gravitational quantum switch introduced in the previous section, as illustrated in figure~\ref{fig.BellITO}(B). In this scenario, the joint state of systems $M$, $S_1$ and $S_2$ after the agents' operations is given by:
\begin{equation}
\frac{\ket{K_{A\prec B}}_M\otimes \mathcal{B}_1\circ\mathcal{A}_1\ket{\psi_1}_{S_1}\otimes \mathcal{B}_2\circ\mathcal{A}_2\ket{\psi_2}_{S_2}
+
\ket{K_{B\prec A}}_M\otimes \mathcal{A}_1\circ\mathcal{B}_1\ket{\psi_1}_{S_1}\otimes \mathcal{A}_2\circ\mathcal{B}_2\ket{\psi_2}_{S_2}}{\sqrt{2}}.
\end{equation}
By performing a measurement of $M$ in the basis $\{\ket{K_{A\prec B}}_M\pm\ket{K_{B\prec A}}_M\}$, the joint state of $S_1$ and $S_2$ becomes
\begin{equation}
\frac{\mathcal{B}_1\circ\mathcal{A}_1\ket{\psi_1}_{S_1}\otimes \mathcal{B}_2\circ\mathcal{A}_2\ket{\psi_2}_{S_2}
\pm
\mathcal{A}_1\circ\mathcal{B}_1\ket{\psi_1}_{S_1}\otimes \mathcal{A}_2\circ\mathcal{B}_2\ket{\psi_2}_{S_2}}{\sqrt{2}}.
\end{equation}

Let us say that systems $S_1$ and $S_2$ are spin $1/2$ particles, and that the initial joint state is $\ket{\uparrow_z}_{S_1}\ket{\uparrow_z}_{S_2}$. If the operations applied by Alice and Bob are given by
\begin{equation}
\mathcal{A}_1=\mathcal{A}_2=\frac{\mathds{1}+\sigma_x}{\sqrt{2}}
\quad \text{and} \quad
\mathcal{B}_1=\mathcal{B}_2=\sigma_z,
\end{equation}
we then obtain one of the final states
\begin{equation}
\frac{\ket{\uparrow_x}_{S_1}\ket{\uparrow_x}_{S_2}\pm\ket{\downarrow_x}_{S_1}\ket{\downarrow_x}_{S_2}}{\sqrt{2}}.
\end{equation}
Note that, regardless of the value that $w$ takes, we have an entangled state that violates a Bell inequality.

It is reasonable to consider that, in a scenario like this, we could guarantee that hypotheses {\it (i)}, {\it (ii)}, {\it (iv)} and {\it (v)} are valid. Hypotheses {\it (i)} and {\it (ii)} depend on the preparation of the experiment. Note that with this, Bell's theorem for temporal order is not device-independent.
Hypotheses {\it (iv)} and {\it (v)} are obeyed if we can guarantee that operations $\mathcal{A}_1$ and $\mathcal{B}_1$ occur in a region that has a space-like separation from the region where $\mathcal{A}_2$ and $\mathcal{B}_2$ occur. Analogously, we can consider that $C_1$, $C_2$, and $D$ occur in regions that have space-like separation between them.
With this, we are left to conclude that {\it (iii)} could not be true, that is, that the order between events is not definite in the protocol we considered.

Just as quantum mechanics is not necessary to formulate a Bell inequality, we do not need hypotheses ~{\it (I)}-{\it (III)}, introduced in previous section, to formulate Bell's theorem for temporal order. However, just as we resort to quantum theory to describe an experiment that violates the inequality, we need these hypotheses to construct a model capable of violating the theorem stated here.

Even though Bell's theorem for temporal order is quite general, it is still theory-dependent~\cite{Kacper}.
Recent works derive a causal inequality, theory-independent and device-independent, that can be violated by a quantum switch coupled to an auxiliary system~\cite{LugtQS,PinzaniQS}.

\subsection{The debate on causal order and its interpretations} \label{SubSecDebate}

As almost always happens in fundamental physics, the research in ICO is surrounded by debates about possible ambiguities and the interpretation of its results. One debate is about the name of the field itself: ``indefinite causal order''.
This name is motivated by the development of formalisms with more general causal structures. Another expression commonly used to denote this area is ``indefinite causality in quantum mechanics''.
However, it is not because a process has indefinite causal order that it will generate some paradox where, for example, a cause occurs after its effect. Therefore, even though some authors call this field ``quantum causality'' or ``indefinite causality'', the way these processes are related to the notion of causality, in the sense of cause and effect, is not so direct. In a quantum switch, for example, it is not operation $\mathcal{A}$ that causes $\mathcal{B}$, or $\mathcal{B}$ that causes $\mathcal{A}$.
Therefore, some authors refer to the quantum switch as a ``superposition of causal order''~\cite{Procopio2015}, while other authors try to avoid even using the word `causal' and prefer to say that the processes have ``indefinite temporal order''~\cite{tbell} or, simply, ``indefinite order of operations''~\cite{PaunkVoji}.

Furthermore, one might be led to interpret the state of the control system as the cause of the order in which operations $\mathcal{A}$ and $\mathcal{B}$ occur. But this interpretation is mistaken. If it were correct, it would also follow that in a gate like the CNot, shown in figure \ref{fig:Circuits1}(C), the control bit is the cause of the Not operation being applied or not to the target system. However, this cannot be true because, in figure~\ref{fig:Circuits1}(F), we have exactly the same operation with the bottom and top bits swapped,  a circuit that is obtained by a mere change of basis from the first one. Therefore, it makes no sense to attribute physical meaning to the control bit as the cause of an effect on the target.

An even more intense debate has occurred over what constitutes a genuine implementation of a quantum switch and what would be just a simulation. Indefinite causal order was proposed in a context where an exotic spacetime was imagined to be necessary in order to implement the processes introduced by the new formalisms. However, some of these processes, in particular the quantum switch, are implemented experimentally without the need for any gravitational effect. As illustrated in figure~\ref{FigOptQS}(B), Alice's and Bob's operations in an optical quantum switch are performed at different pairs of spacetime locations for each state of the control qubit.

According to the definition of what an event is in general relativity, one can conclude that the operation $\mathcal{A}$ performed by Alice (or $\mathcal{B}$ performed by Bob) on the target system would each be associated with two different events, making the protocol immersible in a classical spacetime. Therefore, it is argued that an optical quantum switch must be just a simulation, and not a genuine implementation of a protocol with indefinite causal order~\cite{PaunkVoji}. For this reason, it is usually argued that genuine implementations would occur in gravitational contexts, such as the one we discussed in the previous section.

A counter-argument to this way of counting events is that the correct definition of event is the one proposed in the process matrix formalism, which we introduced in subsection~\ref{SubSecProcICO}. There, an event corresponds to an operation performed only once in a closed laboratory. Since Alice and Bob each perform their operation only once, regardless of the state of the control qubit, each operation corresponds to a single event. Thus, events that appear to be distinct according to general relativity should be understood as the same event in the process matrix formalism~\cite{Procopio2015}. As we discussed in section~\ref{SubSecProcICO}, this definition of events is limited to certain hypotheses and may not be adequate for a general description of nature.

To reinforce that the optical quantum switch is a genuine indefinite causal order process, some authors state that the quantum switch can only be described by a more general mathematical language than that of quantum circuits, such as process matrices~\cite{Procopio2015, Oreshkov2019timedelocalized}. But other works show that it is indeed possible to describe the quantum switch with a circuit~\cite{PaunkVoji, PhysRevA.110.022227, Ormrod2023causalstructurein}. Moreover, some authors go in the direction of generalizing the idea of closed laboratories and the definition of event~\cite{Oreshkov2019timedelocalized,QDiffeo,EventsinLab}, further enriching the discussion.

It is also worth noting that in the optical implementations of a quantum switch, as well as in the gravitational quantum switch proposed in~\cite{tbell}, the closed laboratories hypothesis is not fully satisfied. Recall that the statistics predicted by a process such as the quantum switch are, in theory, obeyed formally for any choice of operations made by Alice and Bob in their laboratories. Let us suppose that Alice and Bob use some internal auxiliary system to decide which operation to apply to the target system. Suppose this auxiliary system is a clock, and that Alice and Bob perform operations $\mathcal{A}'$ and $\mathcal{B}'$, respectively, if the time $t$ measured by their clocks is less than a given value $t^*$. If $t\geq t^*$, Alice and Bob perform operations $\mathcal{A}''$ and $\mathcal{B}''$. Then, the optical experiments would not implement the map given in equation~(\ref{QSunitarias}), but rather the map
\begin{equation}\label{NaoQS}
S_{\mathcal{A}', \mathcal{A}'',\mathcal{B}',\mathcal{B}''}=\ketbra{0}_c\otimes \mathcal{B}''\circ\mathcal{A}'+\ketbra{1}_c\otimes\mathcal{A}''\circ\mathcal{B}'.
\end{equation}

\begin{figure}[t]
   \centering
   \hspace{-0.5cm}\includegraphics[width=0.9\linewidth]{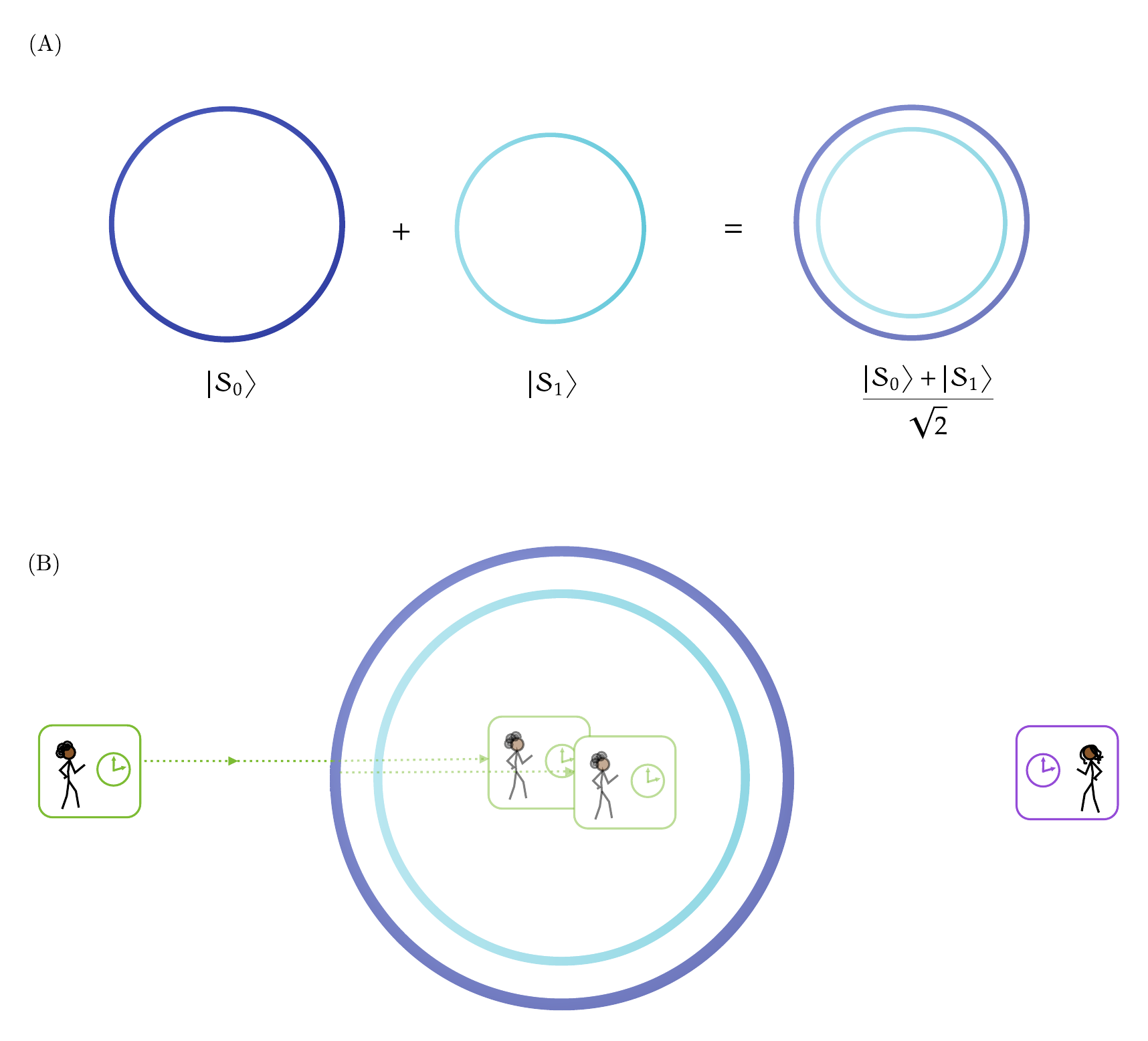}

\caption{Simplified reproduction of the model proposed in~\cite{SMoller2024gravitational} for the realization of a gravitational quantum switch. (A) Illustration of two classical configurations of the same massive system and a superposition of these two configurations. (B) Alice crosses the superposition of spherical shells following geodesics, while Bob remains in the external region.}
\label{fig.Shells}
\end{figure}

In the case of the gravitational quantum switch proposed in~\cite{tbell}, it would still be possible to implement a quantum switch with auxiliary clocks. But if we chose auxiliary systems that measures weight, for example, we would have an analogous problem.

With this, it is worth asking: what is the difference between the optical quantum switch and the gravitational quantum switch? What is so special about gravity for this research field? Does it only serve as motivation to study indefinite causal order, or would there be some property that only non-classical spacetimes would exhibit?

We answer this question in reference~\cite{SMoller2024gravitational}, where the present authors and a collaborator investigated the quantum switch in conditions that only a non-classical gravitational field could offer. In this proposal, we study a model where one of the agents travels along geodesics in a quantum spacetime, in what can be understood as a delocalized Einstein's elevator, illustrated in figure~\ref{fig.Shells}.

This spacetime is generated by a massive body in a superposition of two configurations: a larger spherical shell and a smaller one\footnote{For a better understanding, we present here a simplified version of the spacetime and the protocol introduced in~\cite{SMoller2024gravitational}.}, as illustrated in figure~\ref{fig.Shells}(A). The spacetime in the region external to the two shells generated by the two configurations is the same, therefore classical observables in these regions will always coincide. Thus, the spacetime in this external region can be understood as a classical spacetime. While Bob remains in the external region, Alice crosses the spherical shells, as illustrated in figure~\ref{fig.Shells}(B).

Alice's position becomes entangled with the state of the mass, and we use this resource to induce the realization of a gravitational quantum switch. The most important property is that, at the end of the protocol, Alice naturally recovers a classical trajectory in the region outside the spherical shells. Furthermore, Alice has no means of verifying which trajectory was taken because, as she moves along a geodesic, no observation made inside her closed laboratory allows her to distinguish the global spacetime branch.

A more detailed development of this model is beyond the scope of this article, but it is worth highlighting that until now, this is the most accurate model for implementing a closed laboratory in a protocol with indefinite causal order. Note, however, that this is a thought experiment, and a real implementation would still be challenging. Nevertheless, adapting this protocol to violate the inequalities in~\cite{PinzaniQS,LugtQS,tbell} would be the most robust way we know so far to witness an indefinite causal structure of spacetime.

\section{Final considerations} \label{SecConc}

In this article, we discussed two of the most modern topics at the interface between quantum theory and gravity: gravitationally induced entanglement and indefinite causal order.
Before introducing them, we reviewed important quantum information tools: the Mach-Zehnder interferometer, the Stern-Gerlach experiment, Bell inequalities and entanglement, and the language of quantum circuits. When we finally ventured into the topics at the interface of quantum theory and general relativity, we emphasized how these tools ground the proposals, helping us explore the potentially quantum nature of gravity.
We recall that our main focus is not the detailed investigation of quantum gravity theories, but rather to understand whether a theory that unifies quantum theory and gravitation needs to consider quantum behavior of gravity or not.

Regarding the topic of gravitationally induced entanglement, we reviewed current proposals to generate entanglement between two particles solely through their gravitational interaction, and we discussed the various arguments about what this phenomenon would imply for the quantization of gravity. While experimental attempts to implement this test are currently being developed, theoretical contributions continue to increase, both enriching the discussion on what this phenomenon means, as well as finding more efficient ways to perform the experiment and proposing variations.

For the topic of indefinite causal order,  we examined the first example of a causal inequality and noted that there exist mathematical formalisms aiming to reconcile features of quantum theory and general relativity which are capable of describing processes with indefinite causal order. We explored in detail the quantum switch, the most studied example so far, and presented some models for its implementation. We saw that there are optical experiments that realize this task, and that there is extensive debate about whether they present genuine indefinite causal order or are merely a simulation. We also discussed how thought experiments employing different types of spacetime superpositions could induce a superposition of the causal structure, potentially enabling tasks with indefinite order.

Although there is still no consensus on the interpretation of these phenomena, such research provides us with a deeper understanding of the interface between quantum theory and gravity. By challenging conceptual and experimental limits, these studies drive the development of increasingly precise measurement technologies, stimulate the formulation of new theoretical models, and strengthen the dialogue between traditionally distinct fields. The perspective is to improve our understanding of phenomena in the regime where there is a tension between quantum theory and general relativity. The methods described here illuminate aspects of these theories that are yet not fully understood, and may pave the way for a unified formulation of Physics.

\section*{Acknowledgments}

The authors thank Prof. Nelson Yokomizo for his support and discussions over the years on the topics covered in this article.
B.S. thanks Dr. Esteban Castro Ruiz for continuous discussions about causality, quantum reference frames and GIE, which contributed significantly to the writing of this work. B.S. acknowledges the knowledge acquired in the discipline “Table-top quantum gravity” taught by Dr. Marios Christoudolou, and thanks the colleagues Ofek Begnyat and Tales Rick Perche for important discussions about GIE.
N.S.M. thanks the colleagues Gilberto F. Borges and S. Arash Ghoreishi, for reading preliminary versions of this text and for the indication of some references.

The authors benefited from the activities of COST Action CA23115: Relativistic Quantum Information, URL https://rqi-cost.org/, funded by COST (European Cooperation in Science and Technology).
N.S.M. thanks for the funding of projects INDORBLE 09I03-03-V04-00679 for excellent researchers R2, DeQHOST APVV-22-0570, and QuaSiModo VEGA 2/0156/22.

\end{document}